\documentclass[prb,twocolumn,superscriptaddress,floatfix]{revtex4-1}
\usepackage[english]{babel}
\pdfoutput=1
\usepackage[pdftex]{graphicx}
\usepackage{amsmath}
\usepackage{amssymb}
\usepackage{epstopdf}
\usepackage{color}
\usepackage[normalem]{ulem}
\newcommand{\im}{\mathrm{i}}
\newcommand{\dif}{\mathrm{d}}

\newcommand{\e}{\mathrm{e}}
\newcommand{\h}{\mathrm{h}}
\newcommand{\cb}{\mathrm{c}}
\newcommand{\vb}{\mathrm{v}}
\newcommand{\mrd}{\mbox{\hspace*{0.6mm}}}

\usepackage[pdftex]{hyperref}
\begin{document}
\title{Dynamics of exciton formation and relaxation in photoexcited semiconductors}
\author{Veljko Jankovi\'c} \email{veljko.jankovic@ipb.ac.rs}
\affiliation{Scientific Computing Laboratory, Institute of Physics Belgrade, University of Belgrade, 
Pregrevica 118, 11080 Belgrade, Serbia}
\author{Nenad Vukmirovi\'c}\email{nenad.vukmirovic@ipb.ac.rs}
\affiliation{Scientific Computing Laboratory, Institute of Physics Belgrade, University of Belgrade, 
Pregrevica 118, 11080 Belgrade, Serbia}

\begin{abstract} 
We investigate the dynamics
of the exciton formation and relaxation
on a picosecond time scale following a pulsed photoexcitation of a semiconductor. The study is conducted in the
framework of the density matrix theory complemented with the dynamics controlled truncation scheme.
We truncate the phonon branch of the resulting hierarchy of equations and propose the form of coupling
among single--phonon--assisted and higher--order phonon--assisted density matrices so as to ensure
the energy and particle--number conservation in a closed system.
Time scales relevant for the exciton formation and relaxation processes are determined from
numerical investigations performed on a one--dimensional model for the
values of model parameters representative of a typical organic and inorganic semiconductor.
The exciton dynamics is examined for different values of central frequency of the exciting field, temperature, and microscopic model
parameters, such as the strengths of carrier--carrier and carrier--phonon couplings.
We find that for typical organic semiconductor parameters, formation of bound excitons occurs on a several--hundred--femtosecond
time scale, while their subsequent relaxation and equilibration takes at least several picoseconds. These time scales are consistent
with recent experimental studies of the exciton formation and relaxation in conjugated polymer--based materials.
\end{abstract}

\maketitle{}
\section{Introduction}
The continual and ever--increasing demand for economic and efficient ways of utilizing solar energy drives a huge part of current research activities.
In particular, organic solar cells have developed rapidly in the past decade and have become promising candidates for economically viable large--scale
power generation due to their flexibility, cost effectiveness,
relatively simple fabrication techniques, and mass production.~\cite{tress2014organic,RepProgPhys.73.096401}
Processes upon which the operation of solar cells is based are the light absorption in a semiconducting material and the subsequent conversion of photons
into mobile charge carriers that produce an electric current.~\cite{AcChemRes.42.1691,ChemRev.110.6736}
An optical excitation of a semiconductor creates an exciton, i.e., an electron--hole pair
in which Coulomb attraction between oppositely charged electron and hole prevents their separation.
In a conventional inorganic semiconductor, relatively weak Coulomb interaction (primarily due to large dielectric constant)
results in the exciton binding energy of the order of 10 meV.~\cite{Knoxbook,singhcontribution,RepProgPhys.67.433}
Thus, thermal excitations are likely to split the exciton in an electron and a hole.
On the other hand, in a typical organic semiconductor, the attraction between an electron and a hole is much stronger (mainly due to low dielectric constant),
the exciton binding energy being of the order of or larger than 500 meV.~\cite{BaesslerKoehler,AcChemRes.42.1691} Therefore, while optical absorption in an inorganic semiconductor results in almost
immediate generation of free charges, in an organic semiconductor it leads to formation of tightly bound electron--hole pairs,
which should be separated in order to generate current.~\cite{AcChemRes.42.1691,tress2014organic,ChemRev.110.6736}
This last conclusion has an enormous impact on the design and geometry of organic photovoltaic devices.

Photoexcitation of a semiconductor creates electron--hole pairs in a highly nonequilibrium state.
Apart from the Coulomb interaction, which primarily induces correlations,
the carrier--phonon interaction is also vital for a thorough understanding of nonequilibrium processes taking place in photoexcited semiconductors.
Theoretical approaches for treating these processes are most often based on
the density matrix theory~\cite{kuhncontribution,RevModPhys.74.895}
or the nonequilibrium Green's functions formalism.~\cite{Haug}
Density matrix theory has become the preferred technique
in the treatment of experiments with ultrashort pulses,
since it deals with quantities that depend on one time argument and are directly related to observables.

Previous theoretical studies of the exciton formation process after an ultrafast optical excitation of a semiconductor
were typically focused on inorganic semiconductors. Early studies were conducted in the framework of
the semiclassical Boltzmann approach.~\cite{PhysRevB.54.4660,PhysRevB.55.R16049}
The fully microscopic and quantum theory for the interacting system of electrons, holes, photons, and phonons,
capable of treating a wide variety of optical and excitonic effects after an ultrafast optical excitation of a semiconductor, was elaborated
in Refs.~\onlinecite{Kira2006155,Kirabook,PhysRevLett.87.176401,PSSB:PSSB195,PhysRevB.67.155113}.
On the other hand, the exciton formation from an initial state of two opposite charges in organic
semiconductors was typically modeled by simulating the time evolution of empirical Hamiltonians applied
to small systems, where the effects of the lattice are not included or are treated classically.~\cite{PhysRevB.62.11473,PhysRevB.67.045109}

The main aim of this work was to investigate the dynamics of exciton formation on short (up to several ps) time scale.
This time scale is of particular relevance for the operation of organic solar cells, since it has been well established that the exciton separation
at the interface of donor and acceptor materials occurs on a subpicosecond time scale.~\cite{sci258-1474,jpcc112-4350}
However, the details of the exciton formation and separation process and the factors that determine its efficiency are still not well understood.
In recent years, significant insights have been obtained from subpicosecond time--resolved
experiments performed both on neat materials~\cite{jacs132-17459,jpcc115-9726}
and blends.~\cite{afm22-1116,nmat12-29,science335-1340,nmat12-66,science343-512}
The results of all these experiments highlight the importance of nonequilibrium nature of excitons formed after photoexcitation.

In our study, we employ the Hamiltonian which includes all relevant physical effects in the system: electronic coupling which leads to band formation,
electron--hole interaction which causes exciton formation, electron--phonon interaction that leads to relaxation,
and the interaction with external electromagnetic field.
We do not, however, include the effects of stimulated emission which lead to radiative recombination of excitons,
since we are interested in the exciton dynamics on a short time scale, where these effects are negligible.
From the time evolution of relevant quantities, we identify the time scale of the processes of formation of free charges and bound excitons and
their subsequent relaxation. Rather than focusing on the details of one particular material system,
we have chosen a Hamiltonian whose parameters can be easily varied so that we can identify the influence of different physical effects on relevant time scales.
The study is conducted in the framework of the density matrix formalism combined with the so--called dynamics controlled truncation (DCT) scheme,
firstly developed in 1994 by Axt and Stahl.~\cite{ZPhysB.93.195,ZPhysB.93.205}
This method is particularly suited for a system described by a pair--conserving Hamiltonian which is initially unexcited
and was successfully applied to study the dynamics of exciton formation for near--band--gap excitations and low--excitation
densities.~\cite{PhysRevB.65.035303,siantidis2002theory,siantidis2002formation}
Here, we truncate the phonon branch of the hierarchy so as to ensure that the resulting equations are compatible with the energy and particle--number conservation in a closed system. Furthermore, we propose the form of coupling between single--phonon--assisted and higher--order phonon--assisted
electronic density matrices which is compatible with the energy conservation in a closed system.

The paper is organized as follows. In Sec.~\ref{Sec:std_dct} the general form of the Hamiltonian,
along with the equations which describe the exciton formation process, is presented.
Section~\ref{Sec:2b1d} is devoted to the results of our numerical investigations of the exciton formation process
which are carried out on a one--dimensional model system.
The discussion of our results in light of recent experimental investigations of ultrafast exciton dynamics
is presented in Sec.~\ref{Sec:discussion}, whereas concluding remarks are given in Sec.~\ref{Sec:conclusion}. 

\section{Theoretical framework}\label{Sec:std_dct}
We use the standard two--band semiconductor model
which takes into account the interaction of carriers with the
external electromagnetic field
applied to the semiconductor, as well as carrier--carrier and carrier--phonon interactions.
We will work in the electron--hole picture which is particularly suited for
describing the effects which arise after the optical excitation of an initially unexcited semiconductor.
Notation from Ref.~\onlinecite{RevModPhys.70.145} will be used.
The Hamiltonian has the form
\begin{equation}
 \label{Eq:std_sc_ham}
 H=H_{\mathrm{c}}+H_{\mathrm{ph}}+H_{\mathrm{c-ph}}+H_{\mathrm{c-f}},
\end{equation}
where $H_\mathrm{c}$ describes interacting carriers
\begin{equation}
\label{Eq:h_mat}
 \begin{split}
 H_{\mathrm{c}}&=\sum_{q\in\mathrm{CB}}\epsilon^{\cb}_q c_q^\dagger c_q-
 \sum_{q\in\mathrm{VB}}\epsilon^{\vb}_q d_q^\dagger d_q\\
 &+\frac{1}{2}\sum_{pqkl\in\mathrm{CB}} V^{\cb\cb\cb\cb}_{pqkl} c_p^\dagger c_k^\dagger c_l c_q
 +\frac{1}{2}\sum_{pqkl\in\mathrm{VB}} V^{\vb\vb\vb\vb}_{pqkl} d_q^\dagger d_l^\dagger  d_k  d_p\\
 &+\sum_{\substack{pq\in\mathrm{VB}\\kl\in\mathrm{CB}}}\left(V^{\vb\cb\cb\vb}_{plkq}-V^{\vb\vb\cb\cb}_{pqkl}\right)
 c_k^\dagger d_q^\dagger d_p c_l,
 \end{split}
\end{equation}
\begin{equation}
 \label{Eq:h_p}
H_{\mathrm{ph}}=\sum_{\mu}\hbar\omega_\mu b_\mu^\dagger b_\mu
\end{equation}
is the free--phonon Hamiltonian, $H_{\mathrm{c-ph}}$ describes the carrier--phonon interaction
\begin{equation}
 \label{Eq:h_ep}
 \begin{split}
 {H}_\mathrm{c-ph}&=\sum_{\substack{pq\in\mathrm{CB}\\\mu}}\left(\gamma^\mu_{pq} c_p^\dagger c_q b_\mu^\dagger+
 \gamma^{\mu*}_{pq} c_q^\dagger c_p b_\mu\right)\\
 &-\sum_{\substack{pq\in\mathrm{VB}\\\mu}}\left(\gamma^\mu_{pq} d_q^\dagger d_p b_\mu^\dagger+
 \gamma^{\mu*}_{pq} d_p^\dagger d_q b_\mu\right),
 \end{split}
\end{equation}
whereas the coupling to the optical field is given by
\begin{equation}
\label{Eq:h_opt}
 {H}_{\mathrm{c-f}}=
 -\mathbf{E}(t)\left(\sum_{\substack{p\in\mathrm{VB}\\q\in\mathrm{CB}}}{\mathbf{M}}^{\vb\cb}_{pq}d_p c_q
+\sum_{\substack{p\in\mathrm{CB}\\q\in\mathrm{VB}}}{\mathbf{M}}^{\cb\vb}_{pq}c_p^\dagger d_q^\dagger\right).
\end{equation}

Fermi operators $c_q^\dagger$ ($c_q$) create (annihilate) an electron of energy $\epsilon^\cb_q$ in the single--particle state $q$
in the conduction band,
while Fermi operators $d_q^\dagger$ ($d_q$) create (annihilate) a hole of energy $-\epsilon^\vb_q$ in the single--particle state $q$ in the valence band.
Matrix elements of the Coulomb interaction potential $V(\mathbf x-\mathbf y)$
are defined as
\begin{equation}
\begin{split}
 &V^{\lambda_p\lambda_q\lambda_k\lambda_l}_{pqkl}=\\
&=\int\dif\mathbf{x}\mrd
\dif\mathbf{y}\mrd\phi^{\lambda_p*}_p(\mathbf x)\phi^{\lambda_q}_q(\mathbf x)
V(\mathbf x-\mathbf y)\phi^{\lambda_k*}_k(\mathbf y)\phi^{\lambda_l}_l(\mathbf y),
\end{split}
\end{equation}
where $\phi^{\lambda_p}_p(\mathbf x)$ are single--particle eigenfunctions for an electron in the state $p$ and in the band $\lambda_p$.
Bose operators $b_\mu^\dagger$ ($b_\mu$) create (annihilate) a phonon in mode $\mu$, while $\gamma^\mu_{pq}$ are
carrier--phonon matrix elements. We neglect intraband contributions to the carrier--field interaction and retain only
interband dipole matrix elements.

We note that the Hamiltonian of interacting carriers [Eq.~\eqref{Eq:h_mat}] includes the limiting cases of Wannier and Frenkel excitons.
Namely, when single--particle eigenfunctions are of the Bloch form labeled by a wave vector $\mathbf{k}$,
then under suitable approximations, described e.g.,
in Ref.~\onlinecite{Hanamura1977209}, we obtain the Hamiltonian describing the limiting case of Wannier excitons.
On the other hand, if single--particle
eigenfunctions are taken to be atomic states labeled by a position vector $\mathbf{R}$, then using
approximations that exploit localization properties of this basis set
the Hamiltonian appropriate for the limiting case of Frenkel excitons is obtained.~\cite{PhysRevB.77.085206}

We study the dynamics of exciton formation in photoexcited semiconductors in the framework of the density matrix
theory. Differential equations for dynamic variables are formed and, due to the many--body nature of the problem,
an infinite hierarchy of differential equations is obtained. The main approximation is then the truncation
of the hierarchy, which can be based upon different physical pictures.
The Hamiltonian defined by Eqs.~\eqref{Eq:std_sc_ham}$-$\eqref{Eq:h_opt} has the property that only the interaction
with the optical field can change the number of pair excitations. The DCT scheme relies
upon the aforementioned property of the Hamiltonian and classifies higher--order density matrices according to their
leading order in the optical field.~\cite{ZPhysB.93.195,PhysRevB.53.7244,RevModPhys.70.145}
Namely, when the system is initially in the ground state represented by the vacuum of electron--hole pairs,
the expectation value of the normal--ordered product of $n_\e$ electron operators $c^\dagger$ and $c$,
$n_\h$ hole operators $d^\dagger$ and $d$ and an arbitrary number of phonon operators $b^\dagger$ and $b$ is at least of
the order $m=\max\{n_\e,n_\h\}$ in the applied field.
Therefore, higher--order density matrices are also of higher order in the
optical field and only a finite number of electronic density matrices contributes to the optical response at any given
order in the optical field. The DCT scheme truncates only the electronic branch of the hierarchy and can
be used along with any strategy to deal with the phonon--assisted branch of the hierarchy.~\cite{RepProgPhys.67.433}
We limit ourselves to the case of weak optical field and low carrier densities, in which it is justified to neglect biexcitonic
effects and keep only contributions up to the second order in the optical field.
In Refs.~\onlinecite{PhysRevB.65.035303,RevModPhys.70.145} a reduced treatment of the phonon branch
of the hierarchy, which can be combined with the DCT scheme for the electronic branch of the hierarchy,
was presented. This treatment includes correlation expansion for phonon--assisted variables combined with the Markov
approximation. As a result, phonon--assisted variables are eliminated from the formalism and only two types of
electronic density matrices remain. These are
the interband transition amplitude (excitonic amplitude)
\begin{equation}
 \label{Eq:def_ex_amp}
Y_{ab}=\langle d_a c_b\rangle
\end{equation}
and the electron--hole pair density (excitonic population)
\begin{equation}
 \label{Eq:def_ex_pop}
N_{abcd}=
 \langle c_a^\dagger d_b^\dagger d_c c_d\rangle.
\end{equation}

In this study, we adopt a different strategy for dealing with the phonon--assisted density matrices.
In order to facilitate the truncation of the phonon--assisted branch of the hierarchy, the following generating functions
for the phonon--assisted electronic density matrices are defined:~\cite{RevModPhys.70.145}
\begin{equation}
 \label{Eq:def_ex_amp_ph}
Y_{ab}^{\alpha\beta}=\langle d_a c_b\hat{F}^{\alpha\beta}\rangle,
\end{equation}
\begin{equation}
 \label{Eq:def_ex_pop_ph}
N_{abcd}^{\alpha\beta}=
 \langle c_a^\dagger d_b^\dagger d_c c_d\hat{F}^{\alpha\beta}\rangle,
\end{equation}
\begin{equation}
\label{Eq:def_gen_ph}
 F^{\alpha\beta}=\langle\hat F^{\alpha\beta}\rangle=\left\langle\exp\left(\sum_\rho\alpha_\rho b_\rho^\dagger\right)
 \exp\left(\sum_\rho\beta_\rho b_\rho\right)\right\rangle,
\end{equation}
where $\{\alpha_\rho\}$ and $\{\beta_\rho\}$ are arbitrary sets of real parameters.
As a consequence of the generating--function property,
all phonon--assisted electronic density matrices can be obtained as derivatives of these functions
taken at $\alpha_\mu=\beta_\mu=0$.
The electron and hole populations and correlations $\langle c_a^\dagger c_b\rangle$
and $\langle d_a^\dagger d_b\rangle$, as well as their phonon--assisted
counterparts, do not have to be considered as independent variables in the formalism
since they can be eliminated in favor of $N$ by identities (contraction identities) that are exact within
the second--order treatment.~\cite{RevModPhys.70.145,PhysRevB.53.7244}
The differential equations for
variables $Y_{ab}^{\alpha\beta}$ and $N_{abcd}^{\alpha\beta}$ are given in Appendix~\ref{App:eom}.

The most general form of an electron--hole pair state is~\cite{Hanamura1977209}
\begin{equation}
 |p\rangle=\sum_{\substack{a\in\mathrm{VB}\\b\in\mathrm{CB}}}\psi_{ab} c_b^\dagger d_a^\dagger|0\rangle,
\end{equation}
where $|0\rangle$ represents the state in which the conduction band is completely empty and the valence band is completely filled.
The excitonic basis is defined by the eigenvalue problem $H_\cb|p\rangle=E|p\rangle$ which can be transformed into equations for amplitudes
$\psi_{ab}$:
 \begin{equation}
\label{Eq:def_exc_bas}
 (\epsilon^\cb_b-\epsilon^\vb_a)\psi_{ab}^x+\sum_{\substack{p\in\mathrm{VB}\\q\in\mathrm{CB}}}
 \left(V^{\vb\cb\cb\vb}_{pqba}-V^{\vb\vb\cb\cb}_{pabq}\right)\psi_{pq}^x=\hbar\omega_x\psi_{ab}^x.
\end{equation}
The excitonic basis is orthonormal
\begin{equation}
 \label{Eq:exc_bas_orth}
\sum_{\substack{a\in\mathrm{VB}\\b\in\mathrm{CB}}}\psi_{ab}^{\bar x*}\psi_{ab}^x=\delta_{x\bar x}.
\end{equation}
We perform all calculations in the excitonic basis and expand all density matrices in the excitonic basis, for example
\begin{equation}
\label{Eq:y_ex_bas}
 Y_{ab}=\sum_x \psi^x_{ab}\mrd y_x,
\end{equation}
\begin{equation}
\label{Eq:n_ex_bas}
 N_{abcd}=\sum_{\bar x x}\psi^{\bar x*}_{ba}\psi^x_{cd}\mrd n_{\bar x x},
\end{equation}
and similarly for the corresponding phonon--assisted electronic density matrices; in the case of single--phonon assistance,
the explicit definitions are
\begin{subequations}
 \begin{eqnarray}
  Y_{ab\mu^+}\equiv\langle d_a c_b b_\mu^\dagger\rangle&=&\sum_x \psi^x_{ab} y_{x\mu^+},
\\
  N_{abcd\mu^+}\equiv\langle c_a^\dagger d_b^\dagger d_c c_d b_\mu^\dagger\rangle&=&
\sum_{\bar x x}\psi^{\bar x*}_{ba}\psi^x_{cd} n_{\bar x x\mu^+}.
 \end{eqnarray}
\end{subequations}

The creation operator for the exciton in the state $x$ can be defined as
\begin{equation}
 \label{Eq:def_exc_creat}
  X_x^\dagger=\sum_{\substack{a\in\mathrm{CB}\\b\in\mathrm{VB}}}\psi^x_{ba} c_a^\dagger d_b^\dagger.
\end{equation}
The number of excitons in the state $x$, after performing the decoupling
(which is exact up to the second order in the optical field)
$\langle c_a^\dagger d_b^\dagger d_c c_d\rangle=\langle c_a^\dagger d_b^\dagger\rangle
\langle d_c c_d \rangle+\delta\langle c_a^\dagger d_b^\dagger d_c c_d\rangle$,
where $\delta\langle c_a^\dagger d_b^\dagger d_c c_d\rangle$ stands for the correlated part of the electron--hole pair density,
can be expressed as the sum
\begin{equation}
\label{Eq:num_exc_x}
 \langle X_x^\dagger X_x\rangle=|y_x|^2+\bar n_{xx},
\end{equation}
where $\bar n_{\bar x x}=n_{\bar x x}-y_{\bar x}^* y_x$.
The first term in Eq.~\eqref{Eq:num_exc_x} describes the so--called coherent excitons,
whereas the second term describes the incoherent excitons. Namely, an optical excitation
of a semiconductor firstly induces
single--particle excitations in form of 
optical polarizations and carrier densities. Optical polarizations decay very fast due to
various scattering mechanisms present.~\cite{Kirabook}
Therefore, their squared moduli, which are usually
referred to as coherent excitonic populations,~\cite{PhysRevB.65.035303} do not provide information about the true excitonic populations,
which are the consequence of Coulomb--induced correlations between electrons and holes and which
typically exist in the system for a long time after the decay of optical polarizations.~\cite{RepProgPhys.67.433}
In order to describe true excitons, which are atomlike complexes of electrons and holes
bound by the Coulomb attraction, we have to consider two--particle correlations between them,
and not single--particle quantities.~\cite{Kirabook}
The last conclusion justifies identification of the term $\delta\langle c_a^\dagger d_b^\dagger d_c c_d\rangle$
with the incoherent excitonic populations.

The dynamic equations for the relevant variables should be compatible with the energy conservation in a
system without external fields.
Our system, however, interacts with external optical field, but, since we consider
a pulsed excitation, the energy of the system should be conserved after the field has vanished.
The total energy of the system, i.e., the expectation value of the Hamiltonian $\langle H\rangle$
defined in Eqs.~\eqref{Eq:std_sc_ham}--\eqref{Eq:h_opt}, is expressed as
\begin{equation}
\label{Eq:tot_ene}
 \mathcal{E}=\mathcal{E}_\mathrm{c}+\mathcal{E}_\mathrm{ph}+\mathcal{E}_\mathrm{c-ph}+\mathcal{E}_\mathrm{c-f},
\end{equation}
where the carrier energy is
\begin{equation}
\label{Eq:carr_ene}
 \mathcal{E}_\mathrm{c}=\sum_{x}\hbar\omega_x\mrd n_{xx},
\end{equation}
the phonon energy is
\begin{equation}
\label{Eq:phon_ene}
 \mathcal{E}_\mathrm{ph}=\sum_\mu\hbar\omega_\mu\mrd\langle b_\mu^\dagger b_\mu\rangle,
\end{equation}
the carrier--phonon interaction energy is
\begin{equation}
\label{Eq:carr_phon_ene}
 \mathcal{E}_\mathrm{c-ph}=2\sum_{\bar x x\mu}
 \mathrm{Re}\{\Gamma^\mu_{\bar x x}n_{\bar x x\mu^+}\},
\end{equation}
and the carrier--field interaction energy is
\begin{equation}
\label{Eq:carr_fi_ene}
 \mathcal{E}_\mathrm{c-f}=-\mathbf{E}(t)\sum_x\left(\mathbf{M}_x^* y_x+y_x^*\mathbf{M}_x\right).
\end{equation}
In Eqs.~\eqref{Eq:tot_ene}--\eqref{Eq:carr_fi_ene} we have kept only contributions up to the second order in the external
field and transferred to the excitonic basis.
We also introduce excitonic dipole matrix elements
\begin{equation}
 \mathbf{M}_x=\sum_{\substack{a\in\mathrm{VB}\\b\in\mathrm{CB}}}\psi_{ab}^{x*}\mathbf{M}^{\cb\vb}_{ba},
\end{equation}
as well as matrix elements of the carrier--phonon interaction in the excitonic basis which describe the coupling
to the phonon mode $\mu$:
\begin{equation}
\label{Eq:exc_ph_cc}
 \Gamma^\mu_{xx'}=\sum_{\substack{a\in\mathrm{VB}\\b\in\mathrm{CB}}}\psi_{ab}^{x*}\left(\sum_{k\in\mathrm{CB}}\gamma^\mu_{bk}\psi_{ak}^{x'}-
 \sum_{k\in\mathrm{VB}}\gamma^\mu_{ka}\psi_{kb}^{x'}\right).
\end{equation}
Within previous approaches to solving the hierarchy of equations obtained after performing the DCT scheme,
single--phonon--assisted density matrices $n_{\bar x x\mu^+}$, which appear in Eq.~\eqref{Eq:carr_phon_ene}, were not
explicitly taken into account, but the respective differential equations were solved
in the Markov and adiabatic approximations. However, it can be shown that the total energy under these approximations is not exactly
conserved after the external field has vanished. In order to satisfy the energy conservation, we retain density matrices
$n_{\bar x x\mu^+}$ as independent dynamic variables in the formalism.  

The dynamics should also conserve the particle number after the external field
has vanished, since all the other terms in the Hamiltonian given by Eqs.~\eqref{Eq:std_sc_ham}--\eqref{Eq:h_opt}
commute with the total particle--number operator. The number of electrons (and also the number of holes,
since carriers are generated in pairs in this model), with accuracy up to
the second order in the external field, is given as
\begin{equation}
\label{Eq:def_tot_num}
 N_\mathrm{tot}=N_\e=N_\h=\sum_{x} n_{xx}.
\end{equation}

The equations for the purely electronic relevant variables and phonon distribution function are
\begin{equation}
\label{Eq:dif_y_x}
\begin{split}
 \partial_t y_x&=-\im\omega_x y_x-\frac{1}{\im\hbar}\mathbf{E}(t){\mathbf M}_x\\
&+\frac{1}{\im\hbar}\sum_{\mu x'}\Gamma^\mu_{xx'}\mrd y_{x'\mu^+}+
\frac{1}{\im\hbar}\sum_{\mu x'}\Gamma^{\mu*}_{x'x}\mrd y_{x'\mu^-},
\end{split}
\end{equation}
\begin{equation}
\label{Eq:dif_n_barx_x}
\begin{split}
 \partial_t n_{\bar x x}&=-\im(\omega_x-\omega_{\bar x})n_{\bar x x}-\frac{1}{\im\hbar}\mathbf{E}(t)\left(
 y_{\bar x}^*{\mathbf{M}}_x-{\mathbf{M}}_{\bar x}^* y_x\right)\\
 &+\frac{1}{\im\hbar}\sum_{\mu x'}\Gamma^\mu_{xx'}n_{\bar x x'\mu^+}-
 \frac{1}{\im\hbar}\sum_{\mu \bar x'}\Gamma^\mu_{\bar x'\bar x}n_{\bar x' x\mu^+}\\
 &+\frac{1}{\im\hbar}\sum_{\mu x'}\Gamma^{\mu*}_{x'x}n_{x'\bar x\mu^+}^*-
 \frac{1}{\im\hbar}\sum_{\mu \bar x'}\Gamma^{\mu*}_{\bar x\bar x'}n_{x\bar x'\mu^+}^*,
 \end{split}
\end{equation}
\begin{equation}
\label{Eq:dif_num_ph}
 \partial_t\langle b_\mu^\dagger b_\mu\rangle=\frac{2}{\hbar}\sum_{\bar x x}
 \mathrm{Im}\{\Gamma^\mu_{\bar x x}n_{\bar x x\mu^+}\}.
\end{equation}

Even at this level, without specifying the form of equations for one--phonon--assisted electronic
density matrices, using Eq.~\eqref{Eq:dif_n_barx_x} with vanishing electric field it is easily shown that, in the absence
of external fields, our dynamics conserves the total number of particles.

We will neglect hot--phonon effects and assume that in all the equations
for $y_x$, $n_{\bar x x}$, and their phonon--assisted counterparts
the phonon numbers assume their equilibrium values $n_\mu^{\mathrm{ph}}=(\e^{\beta\hbar\omega_\mu}-1)^{-1}$.
We will, however, retain Eq.~\eqref{Eq:dif_num_ph} in the formalism because it is necessary
to prove the energy conservation. 

In equations for phonon--assisted electronic density matrices we neglect the coupling to the light field, i.e., we neglect
contributions arising from the combined action of the phonon coupling and the interaction with the light field (so--called
cross terms).~\cite{PhysRevB.46.7496,RevModPhys.70.145}
The equations for the electronic density matrices with one--phonon assistance contain
electronic density matrices with
two--phonon assistance, from which
we explicitly separate the factorized part and the correlated part, for example
\begin{equation}
\label{Eq:n_two_ph_ass_fact}
 \langle c_a^\dagger d_b^\dagger d_c c_d b_\mu^\dagger b_\rho\rangle=
\langle c_a^\dagger d_b^\dagger d_c c_d\rangle\delta_{\mu\rho}n_\mu^\mathrm{ph}+
\delta\langle c_a^\dagger d_b^\dagger d_c c_d b_\mu^\dagger b_\rho\rangle,
\end{equation}
\begin{equation}
 \langle d_a c_b b_\mu^\dagger b_\rho\rangle=
 \langle d_a c_b\rangle\delta_{\mu\rho}n_\mu^\mathrm{ph}+\delta\langle d_a c_b b_\mu^\dagger b_\rho\rangle.
\end{equation}
We should bear in mind that the two--phonon--assisted electronic density matrices with two creation
(annihilation) phonon operators,
whose factorized part vanishes, should be considered on this level of truncation of the phonon branch.~\cite{PhysRevB.50.5435}
Further comments on the factorization performed in Eq.~\eqref{Eq:n_two_ph_ass_fact} are given in Appendix~\ref{App:two-ph}. 
The following equations for the electronic density matrices with single--phonon assistance are obtained:
\begin{equation}
\label{Eq:dif_n_barx_x_mu_piu}
\begin{split}
 \partial_t n_{\bar x x \mu^+}&=-\im(\omega_x-\omega_{\bar x}-\omega_\mu)n_{\bar x x \mu^+}\\
 &+\frac{n_\mu^{\mathrm{ph}}}{\im\hbar}\sum_{x'}\Gamma^{\mu*}_{x'x} n_{\bar x x'}\\
 &-\frac{1+n_\mu^{\mathrm{ph}}}{\im\hbar}\sum_{\bar x'} \Gamma^{\mu*}_{\bar x \bar x'}n_{\bar x' x}\\
 &-\frac{1}{\im\hbar}\sum_{\rho \bar x'}\left(\Gamma^{\rho*}_{\bar x\bar x'}\delta n_{\bar x' x\mu^+\rho^-}+
\Gamma^\rho_{\bar x'\bar x}\delta n_{\bar x' x\mu^+\rho^+}\right)\\
 &+\frac{1}{\im\hbar}\sum_{\rho x'}\left(\Gamma^{\rho*}_{x'x}\delta n_{\bar x x'\mu^+\rho^-}+
\Gamma^\rho_{xx'}\delta n_{\bar x x'\mu^+\rho^+}\right),
\end{split}
\end{equation}
\begin{equation}
\label{Eq:dif_y_x_mu_piu}
\begin{split}
 \partial_t y_{x\mu^+}=&-\im(\omega_x-\omega_\mu)\mrd y_{x\mu^+}+
 \frac{n_\mu^{\mathrm{ph}}}{\im\hbar}\sum_{x'}\Gamma^{\mu*}_{x'x}\mrd y_{x'}\\
 &+\frac{1}{\im\hbar}\sum_{\rho x'}\left(\Gamma^\rho_{xx'}\delta y_{x'\mu^+\rho^+}+
 \Gamma^{\rho*}_{x'x}\delta y_{x'\mu^+\rho^-}\right),
\end{split}
\end{equation}
\begin{equation}
\label{Eq:dif_y_x_mu_meno}
\begin{split}
 \partial_t y_{x\mu^-}=&-\im(\omega_x+\omega_\mu)\mrd y_{x\mu^-}+
 \frac{1+n_\mu^{\mathrm{ph}}}{\im\hbar}\sum_{x'}\Gamma^{\mu}_{xx'}\mrd y_{x'}\\
 &+\frac{1}{\im\hbar}\sum_{\rho x'}\left(
 \Gamma^\rho_{xx'}\delta y_{x'\rho^+\mu^-}+\Gamma^{\rho*}_{x'x}\delta y_{x'\rho^-\mu^-}
 \right).
\end{split} 
\end{equation}

The correlated parts of two--phonon--assisted density matrices appearing
in Eqs.~\eqref{Eq:dif_n_barx_x_mu_piu} ($\delta n_{\bar x x\mu^+\rho^-},\delta n_{\bar x x\mu^+\rho^+}$),
~\eqref{Eq:dif_y_x_mu_piu}, and~\eqref{Eq:dif_y_x_mu_meno}
can be obtained solving their respective differential equations,
in which all three--phonon--assisted density matrices
have been appropriately factorized and their correlated
parts have been neglected, in the Markov and adiabatic approximations.
This procedure closes the phonon branch of the hierarchy.
However, the full solution to these equations, when combined with Eq.~\eqref{Eq:dif_n_barx_x_mu_piu},
is cumbersome to evaluate, so further approximations are usually employed.
The most common one is the so--called random phase approximation, which neglects sums over
correlated parts of one--phonon--assisted electronic density matrices
(which are complex quantities)
due to random phases at different
arguments of these density matrices.~\cite{kuhncontribution}
After performing all the discussed approximations, the last two summands in Eq.~\eqref{Eq:dif_n_barx_x_mu_piu},
which represent the rate at which $n_{\bar x x\mu^+}$ changes due to the coupling to electronic density matrices
with higher phonon assistance,
reduce to
\begin{equation}
\label{Eq:result_RPA}
 \left(\partial_t n_{\bar x x\mu^+}\right)_\mathrm{higher}=-\gamma_{\bar x x\mu}n_{\bar x x\mu^+},
\end{equation}
where $\gamma_{\bar x x\mu}$ is given as
\begin{equation}
\label{Eq:def_gamma_barx_x_mu}
 \gamma_{\bar x x\mu}=\frac{1}{2}\left(\Gamma_x+\Gamma_{\bar x}\right),
\end{equation}
\begin{equation}
\label{Eq:attach_def_gamma_barx_x_mu}
\begin{split}
 \Gamma_x=\frac{2\pi}{\hbar}\sum_{\tilde x\rho}
 \left(|\Gamma^\rho_{x\tilde x}|^2\delta(\hbar\omega_x-\hbar\omega_{\tilde x}+\hbar\omega_\rho)n_\rho^\mathrm{ph}
 \right. \\+ \left.
 |\Gamma^\rho_{\tilde x x}|^2\delta(\hbar\omega_x-\hbar\omega_{\tilde x}-\hbar\omega_\rho)(1+n_\rho^\mathrm{ph})
 \right).
\end{split}
\end{equation}
Details of the procedure employed to close the phonon branch of the hierarchy are given in Appendix~\ref{App:two-ph}.

It was recognized that this form of the coupling to higher--order phonon--assisted electronic density matrices is
at variance with the energy conservation.~\cite{PSSB:PSSB2221880139,kuhncontribution,RevModPhys.74.895}
In this work, we will use the following form of the coupling to
higher--order phonon--assisted density matrices:
\begin{equation}
\label{Eq:our_ansatz}
 \left(\partial_t n_{\bar x x\mu}^{(+)}\right)_\mathrm{higher}=
 -\gamma_{\bar x x\mu}n_{\bar x x\mu}^{(+)}+\gamma_{\bar x x\mu} n_{\bar x x\mu}^{(+)*},
\end{equation}
where $\gamma_{\bar x x\mu}$ is, as before,
defined by Eqs.~\eqref{Eq:def_gamma_barx_x_mu} and~\eqref{Eq:attach_def_gamma_barx_x_mu}.
This form of $\left(\partial_t n_{\bar x x\mu}^{(+)}\right)_\mathrm{higher}$
is compatible with the energy conservation, as long as excitonic matrix elements of the
carrier--phonon interaction $\Gamma^\mu_{\bar x x}$ are purely real, which is the case relevant for our numerical
investigation in Sec.~\ref{Sec:2b1d}.
Namely, as is shown in Appendix~\ref{App:ene_con}, the rate at which the total energy changes after the pulse is equal to
the rate at which the carrier--phonon interaction energy changes due to the coupling of the single--phonon--assisted
electronic density matrices
$n_{\bar x x\mu^+}$ to density matrices with higher--order phonon assistance,
\begin{equation}
\begin{split}
 \partial_t\mrd\mathcal{E}&=\left(\partial_t\mrd\mathcal{E}_\mathrm{c-ph}\right)_\mathrm{higher}\\
&=2\sum_{\bar x x\mu}
 \mathrm{Re}\left\{\Gamma^\mu_{\bar x x}\left(\partial_t\mrd n_{\bar x x\mu^+}\right)_\mathrm{higher}\right\}.
\end{split}
\end{equation}
It is then clear that, if all $\Gamma^\mu_{\bar x x}$ are real, the form of $\left(\partial_t\mrd n_{\bar x x\mu^+}\right)_\mathrm{higher}$
given in Eq.~\eqref{Eq:our_ansatz} does not violate the energy conservation.
Furthermore, as $n_{\bar x x\mu^+}$ describes the elementary process in which
an exciton initially in the state $x$ is scattered to the state $\bar x$ emitting the phonon from the mode $\mu$,
the reverse
microscopic process, described by $n_{x \bar x\mu^-}=n_{\bar x x\mu^+}^*$, is also possible, so in the differential equation
for $n_{\bar x x\mu^+}$ the quantity $n_{\bar x x\mu^+}^*$ may appear. In Appendix~\ref{App:ene_con},
we comment on the energy conservation in greater detail.

Similar strategy can be adopted to simplify the coupling to electronic density matrices with higher phonon
assistance in~\eqref{Eq:dif_y_x_mu_piu} and~\eqref{Eq:dif_y_x_mu_meno}, with the final result
\begin{equation}
\label{Eq:result_apps_y}
 \left(\partial_t y_{x\mu}^{(\pm)}\right)_\mathrm{higher}=-\gamma_{x\mu}\mrd y_{x\mu}^{(\pm)},
\end{equation}
where
\begin{equation}
\label{Eq:result_apps_y_c}
 \gamma_{x\mu}=\frac{1}{2}\Gamma_x,
\end{equation}
and $\Gamma_x$ is defined in Eq.~\eqref{Eq:attach_def_gamma_barx_x_mu}.

An alternative route to derive equations for the relevant variables is to rewrite the Hamiltonian given in Eq.~\eqref{Eq:std_sc_ham}
in terms of operators $X_x,X_x^\dagger$ [see Eq.~\eqref{Eq:def_exc_creat}], keeping only contributions whose expectation values are at most
of the second order in the optical field. The result is
\begin{equation}
\begin{split}
 H=&\sum_x \hbar\omega_x X_x^\dagger X_x+\sum_\mu \hbar\omega_\mu b_\mu^\dagger b_\mu\\
  +&\sum_{\mu \bar x x}\left(\Gamma^\mu_{\bar x x}X_{\bar x}^\dagger X_x b_\mu^\dagger+
\Gamma^{\mu*}_{\bar x x}X_{x}^\dagger X_{\bar x} b_\mu\right)\\
-&\mathbf{E}(t)\sum_x\left(\mathbf{M}_x^* X_x+\mathbf{M}_x X_x^\dagger\right).
\end{split}
\end{equation}
The excitonic operators (up to the second order in the optical field) satisfy Bose commutation relations $[X_x,X_{\bar x}^\dagger]=\delta_{x\bar x}.$
In this representation,~\cite{PhysRevB.60.2599} the excitons are treated as noninteracting bosons and
the form of the exciton--phonon interaction is transparent, with exciton--phonon coupling constants $\Gamma^\mu_{\bar x x}$
defined in Eq.~\eqref{Eq:exc_ph_cc}.
\section{One--dimensional semiconductor model and numerical results}\label{Sec:2b1d}
Numerical computations will be carried out on a two--band one--dimensional semiconductor model.
We use a tight--binding model on a one--dimensional lattice
with $N$ sites and lattice spacing $a$ to describe the semiconductor.
Periodic boundary conditions are used.
The Hamiltonian describing interacting carriers is given as
\begin{equation}
\begin{split}
 H_\mathrm{c}&=\sum_{i=0}^{N-1} \epsilon^\mathrm{c}_0\mrd c_i^\dagger c_i-
 \sum_{i=0}^{N-1} J^\mathrm{c}(c_i^\dagger c_{i+1}+c_{i+1}^\dagger c_i)\\
 &-\sum_{i=0}^{N-1} \epsilon^\mathrm{v}_0\mrd d_i^\dagger d_i+
 \sum_{i=0}^{N-1} J^\mathrm{v}(d_i^\dagger d_{i+1}+d_{i+1}^\dagger d_i)\\
 &+\frac{1}{2}\sum_{i,j=0}^{N-1}(c_i^\dagger c_i-d_i^\dagger d_i)V_{ij}(c_j^\dagger c_j-d_j^\dagger d_j).
\end{split} 
\end{equation}
It is assumed that the carrier transfer integrals
$J^\mathrm{c},J^\mathrm{v}$ are non--zero only among nearest--neighbor pairs of sites.
The Coulomb interaction is taken in the lowest monopole--monopole approximation,~\cite{PSSB:PSSB2221240118} and the interaction
potential $V_{ij}$ is taken to be the Ohno potential
\begin{equation}
 V_{ij}=\frac{U}{\sqrt{1+\left(\frac{|i-j|a}{a_0}\right)^2}}.
\end{equation}
$U$ is the on--site carrier--carrier interaction, while $a_0$ is the characteristic length given as
$a_0=e^2/(4\pi\varepsilon_0\varepsilon_r U)$, where $\varepsilon_r$ is the
static relative dielectric constant.
This form of carrier--carrier interaction is an interpolation between the on--site Coulomb interaction $U$
and the ordinary Coulomb potential (in which the static relative dielectric constant is taken)
$e^2/(4\pi\varepsilon_0\varepsilon_r r)$ when $r\to\infty$ (see, e.g., the discussion on the effective electron--hole interaction
in Ref.~\onlinecite{Knoxbook}).
The interaction with phonons is taken to be of the Holstein form, where a charge carrier is locally
and linearly coupled to a dispersionless optical mode
\begin{equation}
\label{Eq:c_ph_model}
 \begin{split}
 H_\mathrm{c-ph}=\sum_{i=0}^{N-1} g^\cb\mrd c_i^\dagger c_i(b_i+b_i^\dagger)-
 \sum_{i=0}^{N-1} g^\vb\mrd d_i^\dagger d_i(b_i+b_i^\dagger),
 \end{split}
\end{equation}
where the free--phonon Hamiltonian is
\begin{equation}
H_\mathrm{ph}=\sum_{i=0}^{N-1}\hbar\omega_\mathrm{ph} b_i^\dagger b_i. 
\end{equation}
Despite the fact that the carrier--phonon interaction in real materials has a more complicated form,
we choose for our numerical investigations its simplest possible form (Eq.~\eqref{Eq:c_ph_model}) capable of
providing the energy relaxation of the electronic subsystem.
The interaction with the electric field is
\begin{equation}
 H_\mathrm{c-f}=-d_\mathrm{cv}E(t)\sum_{i=0}^{N-1}(d_i c_i+c_i^\dagger d_i^\dagger).
\end{equation}
As the system described is translationally symmetric, we can transfer to the momentum space and obtain
the same Hamiltonian as described in Eqs.~\eqref{Eq:std_sc_ham}--\eqref{Eq:h_opt} with the following values
of parameters:
\begin{subequations}
\label{Eq:after_Fourier_all}
 \begin{eqnarray}
  \epsilon^{\cb/\vb}_k=\epsilon^{\cb/\vb}_0-2J^{\cb/\vb}\cos(ka),\\
  \gamma^q_{k_1k_2}=\delta_{k_2,k_1+q}\frac{g^\cb}{\sqrt{N}}\mrd\mathrm{for}\mrd k_1,k_2\in\mathrm{CB},\\
  \gamma^q_{k_1k_2}=\delta_{k_1,k_2+q}\frac{g^\vb}{\sqrt{N}}\mrd\mathrm{for}\mrd k_1,k_2\in\mathrm{VB},\\
  V^{\vb\vb\cb\cb}_{pqkl}=\delta_{k+q,p+l} V_{k-l},\quad V^{\vb\cb\cb\vb}_{plkq}=0.
 \end{eqnarray}
\end{subequations}
The signs of the transfer integrals are $J^\mathrm{c}>0,J^\mathrm{v}<0$. The constant energy
$\epsilon_0^\mathrm{c}>0$, while $\epsilon_0^\mathrm{v}<0$ is chosen so that the maximum of the valence
band is the zero of the energy scale.
$V_{k-l}$ is the Fourier transformation of the Ohno potential and it is computed numerically as
\begin{equation}
 V_k=\frac{1}{N^2}\sum_{i,j=0}^{N-1}V_{ij}\mrd\e^{-\im ka(i-j)}.
\end{equation}

The translational symmetry of our model enables us to solve efficiently the eigenvalue problem~\eqref{Eq:def_exc_bas}
which defines the excitonic basis. Instead of solving eigenvalue problem of a $N^2\times N^2$ matrix,
we can solve $N$ independent eigenvalue problems of matrices of dimension $N\times N$, thus obtaining $N^2$ excitonic
eigenstates and their eigenenergies, which are counted by the center--of--mass wave vector $Q$ and the band index $\nu$.
Thus, the general index of an excitonic state $x$ should be, in all practical calculations, replaced by combination $(Q,\nu)$.
This has the following consequences on the matrix elements in the excitonic basis: dipole matrix elements reduce to
\begin{equation}
  {M}_{(Q\nu)}=\delta_{Q,0}\mrd d_{\cb\vb}\sum_{k_e}\psi_{Q-k_e,k_e}^{(Q\nu)*},
 \end{equation}
whereas carrier--phonon interaction matrix elements reduce to
\begin{equation}
\label{Eq:carr_phon_sel_r}
\begin{split}
 &\Gamma^q_{(Q\nu)(Q'\nu')}=\delta_{Q',Q+q}
\frac{1}{\sqrt N}\sum_{k_e}\psi_{Q-k_e,k_e}^{(Q\nu)*}\\
&\times
\left(
  g^\cb\psi_{Q-k_e,Q'-Q+k_e}^{(Q'\nu')}-
  g^\vb\psi_{Q'-k_e,k_e}^{(Q'\nu')}
  \right).
\end{split} 
\end{equation}

Due to the translational symmetry of our model, only the dynamic variables for which 
the total created wave vector is equal to the total annihilated wave vector will have
nontrivial values in the course of the system's evolution. 
For example, from all density matrices $y_{(Q\nu)}$ only those with $Q=0$ can have
non--zero values.

Our objective is to analyze, in the framework of this relatively simple model, the characteristic time scales
of exciton formation and relaxation in a photoexcited semiconductor, along with the impact that various model parameters
have on these processes. Basic parameters in our model are transfer integrals
$J^\cb,J^\vb$ (which determine bandwidths of the conduction and valence bands),
electron--phonon coupling constants $g^\cb,g^\vb$, the phonon energy $\hbar\omega_\mathrm{ph}$,
the dielectric constant $\varepsilon_r$, and the on--site Coulomb interaction $U$. We will, throughout the computations,
assume for simplicity that $J^\cb=J^\vb=J$ and $g^\cb=g^\vb=g$.

As is well known, the main differences between a typical organic and inorganic semiconductor can be
expressed in terms of bandwidths, dielectric constant and the carrier--phonon interaction strength.
Namely, inorganic semiconductors are
characterized by wide bands and high value of dielectric constant, whereas organic semiconductors
have narrow bands and small value of dielectric constant. The carrier--phonon interaction is
stronger in organic than in inorganic semiconductors.
Having all these facts in mind, we propose two sets of model parameters which assume values typical
of an organic and inorganic semiconductor. Values of our model parameters are adjusted to material parameters
of GaAs for the inorganic case and pentacene for the organic case. Values of carrier--phonon coupling constants
are chosen to correspond to typical values for mobility and/or typical values for the polaron binding energy.

Typical bandwidths in organic semiconductors are $W\sim 500\mrd\mathrm{meV}$,~\cite{BaesslerKoehler} which corresponds to the transfer
integral $J\sim 125\mrd\mathrm{meV}$, whereas inorganic semiconductors usually exhibit
bandwidths of several electronvolts~\cite{BaesslerKoehler} and we take in our calculations the value of the transfer integral
$J=500\mrd\mathrm{meV}$.
In both cases, the lattice constant was fixed to $a=1\mrd\mathrm{nm}$. The dielectric constant
in a typical inorganic semiconductor is of the order of 10 and in the calculations we take the
value of static dielectric constant of GaAs $\varepsilon_r=12.9$. For a representative value of the
dielectric constant in organic semiconductors we take $\varepsilon_r=3.0$.~\cite{BaesslerKoehler,ChemRev.110.6736}
The value of the on--site
Coulomb interaction $U$ is chosen to give the correct order of magnitude for the exciton binding energy,
which is calculated numerically. For the organic parameter set, we set $U=480$ meV, which gives the exciton binding energy
around 320 meV, while for the inorganic parameter set $U=15$ meV and the corresponding exciton binding energy is roughly 10 meV.

The carrier--phonon coupling constants for the inorganic case are estimated from the mobility values.
The mobility of carriers is estimated using the relation $\mu=e\tau/m^*$,
where $\tau$ is the scattering time and $m^*$ is the effective mass of a carrier. For cosine bands considered
in this work, $m^*=\hbar^2/(2|J|a^2)$ in the vicinity of the band extremum.
The scattering time is estimated from the expression for the carrier--phonon inelastic scattering rate based on
the Fermi's golden rule, which around the band extremum $k=0$ assumes the following form 
\begin{equation}
\label{Eq:scatt_time}
 \frac{1}{\tau(k)}=\frac{g^2}{\hbar|J|}
 \frac{n^\mathrm{ph}}{\sqrt{1-\left(\cos(ka)-\frac{\hbar\omega_\mathrm{ph}}{2|J|}\right)^2}},
\end{equation}
where $n^\mathrm{ph}=(\e^{\beta\hbar\omega_\mathrm{ph}}-1)^{-1}$. Therefore, the carrier--phonon coupling constant
in terms of the carrier mobility reads as
\begin{equation}
 g=|J|\sqrt{\frac{2ea^2}{\hbar\mu n^\mathrm{ph}}}
 \left(1-\left(1-\frac{\hbar\omega_\mathrm{ph}}{2|J|}\right)^2\right)^{1/4}.
\end{equation}
Using the value for the electron mobility in GaAs at $300\mrd\mathrm{K}$
$\mu_\e\approx 8500\mrd\mathrm{cm}^2/(\mathrm{Vs})$,~\cite{handbook_params} we obtain $g\approx 25\mrd\mathrm{meV}$.

We can also estimate the carrier--phonon coupling constants from the polaron binding energy. As an estimate of this
quantity, we use the result of the second--order weak--coupling perturbation theory at $T=0$
in the vicinity of the point $k=0$:~\cite{jcp.128.114713}
\begin{equation}
\label{Eq:pol_bind_ene}
 \epsilon_\mathrm{b}^\mathrm{pol}(k)=\frac{g^2}{2|J|}
 \frac{1}{\sqrt{\left(\cos(ka)+\frac{\hbar\omega_\mathrm{ph}}{2|J|}\right)^2-1}}.
\end{equation}
It is known that polaron binding energies in typical inorganic semiconductors are
$\epsilon_\mathrm{b}^\mathrm{pol}\sim 1\mrd\mathrm{meV}$ and we used this fact along with
Eq.~\eqref{Eq:pol_bind_ene} to check our estimate for $g$ from the value of mobility; for
$g\approx 25\mrd\mathrm{meV}$, we obtain $\epsilon_\mathrm{b}^\mathrm{pol}\approx 2\mrd\mathrm{meV}$.
The polaron binding energies in polyacenes lie in the range between $21\mrd\mathrm{meV}$
and $35\mrd\mathrm{meV}$.~\cite{Pope} The value of $g$ in the set of model parameters representative of organic
semiconductors was estimated from the polaron binding energy in pentacene, which is around $20\mrd\mathrm{meV}$.
We obtain that $g\approx 40\mrd\mathrm{meV}$. The values used for the organic/inorganic set of parameters are
listed in Table~\ref{Tab:model_params}.

\begin{table}
 \caption{Model parameters which are representative of a typical organic and inorganic semiconductor. References
from which material parameters are taken are indicated.}
 \label{Tab:model_params}
 \centering
 \begin{tabular}{c c c}
  \hline\hline
  Parameter & Inorganic & Organic\\
  \hline
  $E_g$ (meV) & 1519~\cite{PhysRevB.65.035303} & 2000~\cite{JPCM.27.113204}\\ 
  $J$ (meV) & 500 & 125\\
  $\varepsilon_r$ & 12.9~\cite{PhysRevB.65.035303} & 3.0~\cite{BaesslerKoehler}\\
  $g$ (meV) & 25 & 40\\
  $\hbar\omega_\mathrm{ph}$ (meV) & 36.4~\cite{PhysRevB.65.035303} & 10.0~\cite{PhysRevLett.96.086601,SurfSci.601.3765}\\
  $U$ (meV) & 15 & 480
 \end{tabular}
\end{table}

The form of the electric field is assumed to be a rectangular cosine pulse
\begin{equation}
 E(t)=E_0\cos(\omega_c t)\theta(t+t_0)\theta(t_0-t),
\end{equation}
where $\omega_c$ is the central frequency of the field and $\theta(t)$ is the Heaviside step function.
Time $t_0$ is chosen large enough so that the pulse is
so spectrally narrow that the notion of the central frequency makes sense. On the other hand, the pulse
should be as short as possible, so that after its end we observe the intrinsic dynamics of our system,
the one which is not accompanied by the carrier generation process, but merely shows how initially generated populations
are redistributed among various states. Trying to reconcile the aforementioned requirements, we choose $t_0=250$ fs.
The amplitude of the electric field $E_0$ and the
interband dipole matrix element $d_\mathrm{cv}$ are chosen so that we stay in the low--density regime; particularly,
we choose them so that the corresponding Rabi frequency $\hbar\omega_R=d_\mathrm{cv}E_0$ assumes the value of $0.2$ meV,
which is smaller than any energy scale in our problem and ensures that the excitation is weak.

In order to quantitatively study the process of exciton formation after a pulsed excitation of a semiconductor,
we solved the system of quantum kinetic equations
for electronic density matrices $y_x,n_{\bar x x}$ and their single--phonon--assisted counterparts
[Eqs.~\eqref{Eq:dif_y_x},~\eqref{Eq:dif_n_barx_x},~\eqref{Eq:dif_n_barx_x_mu_piu},~\eqref{Eq:dif_y_x_mu_piu}, and~\eqref{Eq:dif_y_x_mu_meno}
supplemented with Eqs.~\eqref{Eq:result_RPA} and~\eqref{Eq:result_apps_y}]
using the fourth--order Runge--Kutta algorithm.
The computations are performed for the temperature $T=300$ K and the central frequency of the pulse equal to the single--particle gap
($\hbar\omega_c=E_g$).
The exciton is considered bound (unbound) if its energy $\hbar\omega_{(Q\nu)}$ is smaller (larger) than
the smallest single--particle energy difference $\epsilon^\cb_{k_\e}-\epsilon^\vb_{Q-k_\e}$.~\cite{JPCM.27.113204} The equation of
the boundary line which separates bound from unbound pair states reads as
\begin{equation}
\label{Eq:sep}
 \epsilon_\mathrm{sep}(Q)=\epsilon_0^\mathrm{c}-\epsilon_0^\mathrm{v}-
 2\sqrt{(J^{\mathrm{c}})^2+(J^{\mathrm{v}})^2-2J^\mathrm{c}J^\mathrm{v}\cos(Qa)}.
\end{equation}
An unbound exciton may be considered as (quasi)free electron and hole, so this way it is possible
to distinguish between bound excitons
and free carriers.

The pulsed excitation of a semiconductor leads, in the first step, to the generation of coherent electron--hole pairs
that are described in our formalism by the coherent pair amplitudes $y_x$.
The decay of the coherent pair occupation
\begin{equation}
 N_\mathrm{coh}=\sum_x |y_x|^2
\end{equation}
is due to the scattering processes which initiate already during the generation of the pairs and gives a direct measure
of the loss of coherence.~\cite{PhysRevB.65.035303} At the same time, incoherent pair occupations start to grow, driven by
the loss rate of coherent pair occupations.~\cite{RevModPhys.70.145,PhysRevB.65.035303}
In order to quantify the process of exciton formation, we will follow the time dependence of the total number of
incoherent bound excitons  
\begin{equation}
 N_\mathrm{incoh,b}=\sum_{x\in\mathrm{bound}}(n_{xx}-|y_x|^2).
\end{equation}
This quantity represents the number of truly bound electron--hole pairs which exist even after the optical field has vanished
and as such is the direct measure of the efficiency of the exciton formation process. We will, when useful, also consider the number of incoherent
excitons in a particular band $\nu$, $N_{\mathrm{incoh},\nu}$.
The quantities $N_\mathrm{incoh,b}$ and $N_{\mathrm{incoh},\nu}$ will be normalized to the total number of excitons $N_\mathrm{tot}$
defined in Eq.~\eqref{Eq:def_tot_num}.

\subsection{Numerical results: Organic set of parameters}
\label{Subsec:num_org}
\begin{figure}[htbp]
 \begin{center}
  \includegraphics[width=0.4\textwidth]{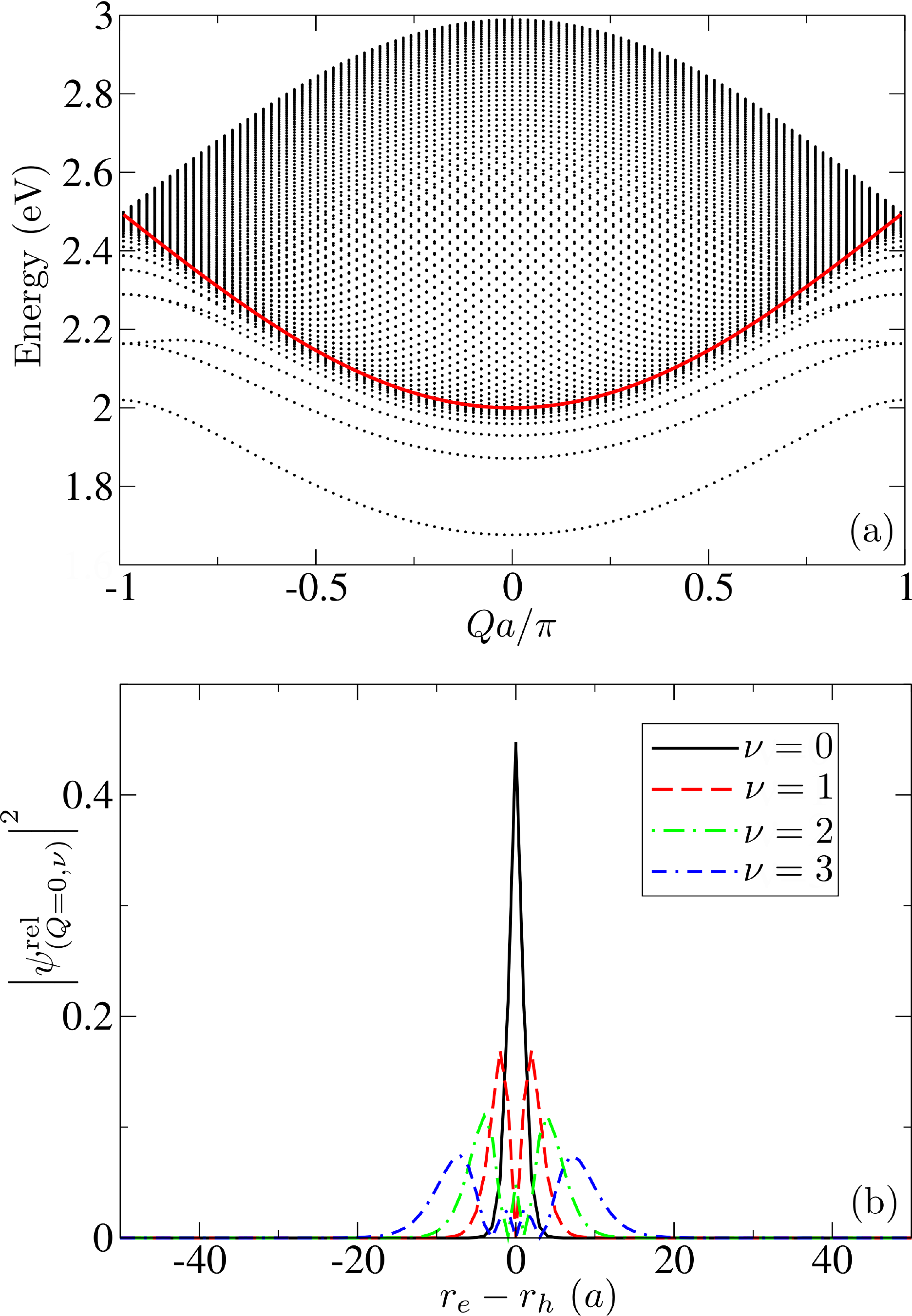}
 \end{center}
 \caption{(Color online)
(a) Excitonic spectrum for the organic set of parameters. Dots represent individual
excitonic states $(Q,\nu)$, while thick red line is the boundary between
bound and unbound excitonic states computed using Eq.~\eqref{Eq:sep}.
(b) Squared modulus of the wave function which describes the relative motion of an electron--hole pair [Eq.~\eqref{Eq:ex_rel_mot}]
 calculated for different states $(Q=0,\nu)$. Mean electron--hole
 separations in these states are $0.7a$ ($\nu=0$), $2.5a$ ($\nu=1$), $4.6a$ ($\nu=2$), and $7.8a$ ($\nu=3$).
Computations are performed for $N=101$.}
 \label{fig:spec_org}
\end{figure}

We start this section by an overview of properties of the excitonic spectrum,
shown in in Fig.~\ref{fig:spec_org}(a),
which will be relevant for further discussions of
the exciton formation process.
The lowest excitonic band is energetically well separated from
the rest of the spectrum, the energy separation
between the minima of the
bands $\nu=0$ and $\nu=1$ being around $200\mrd\mathrm{meV}$, which is much larger than
both the value of $k_BT$ at room temperature and the phonon energy in our model (see Table~\ref{Tab:model_params}).
As a consequence, downward transitions that
end at the lowest excitonic band start almost exclusively from the states on $\nu=1$ band and an exciton, which is
at some instant in a state on the $\nu=0$ band, cannot be scattered to an unbound excitonic state.

We briefly comment on the size of the exciton for these values of model parameters.
From the exciton wave function $\psi^{(Q\nu)}_{Q-k_e,k_e}$ in $k$ space, we can obtain the exciton
wave function in real space performing the Fourier transformation
\begin{equation}
\label{Eq:ex_wf_real}
\begin{split}
 \psi^{(Q\nu)}_{r_e,r_h}&=\sum_{k_e}\e^{\im(Q-k_e)r_h}\e^{\im k_e r_e}\psi^{(Q\nu)}_{Q-k_e,k_e}\\
 &=\e^{\im Q(r_e+r_h)/2}\mrd\sum_{k_e}\e^{-\im(Q-2k_e)(r_e-r_h)/2}\psi^{(Q\nu)}_{Q-k_e,k_e}.
\end{split} 
\end{equation}
The exciton wave function in real space is a product of the plane wave which describes the motion of the center of mass
with the wave vector $Q$ and the wave function of the relative motion of an electron and a hole:
\begin{equation}
\label{Eq:ex_rel_mot}
 \psi_{(Q,\nu)}^\mathrm{rel}=\sum_{k_e}\e^{-\im(Q-2k_e)(r_e-r_h)/2}\psi^{(Q\nu)}_{Q-k_e,k_e}.
\end{equation}
The latter part is
directly related to the exciton size. We calculated squared modulus of the wave function of the relative motion
of a pair for states $(Q=0,\nu)$ in various bands. The result is shown in Fig.~\ref{fig:spec_org}(b).
It is clearly seen that an electron and a hole are tightly bound in these states and their relative separations are of
the order of lattice constant, which is the typical value for the exciton radius in organic semiconductors. We point out
that this does not mean that an exciton is localized; due to the translational symmetry of our system, it is
delocalized over the whole lattice, as described by the plane--wave factor in the total wave function of a pair.
Moreover, we note that the system size $N=101$ is large enough for the results to be numerically accurate, as it is much larger
than the typical size of the exciton in a bound state.

The impact that different parameters have
on the exciton formation process is studied by changing one parameter, at the same time fixing
the values of all the other parameters to the previously mentioned ones. We performed all the computations for a limited
number of lowest excitonic bands, which crucially depends on the central frequency $\omega_c$ of the excitation.
For the given excitation, we took into account all the bands whose minima lie below $\hbar\omega_c+\alpha k_BT$,
where $\alpha\sim 5$ is a numerical constant.

\begin{figure}[htbp]
 \begin{center}
  \includegraphics[width=0.4\textwidth]{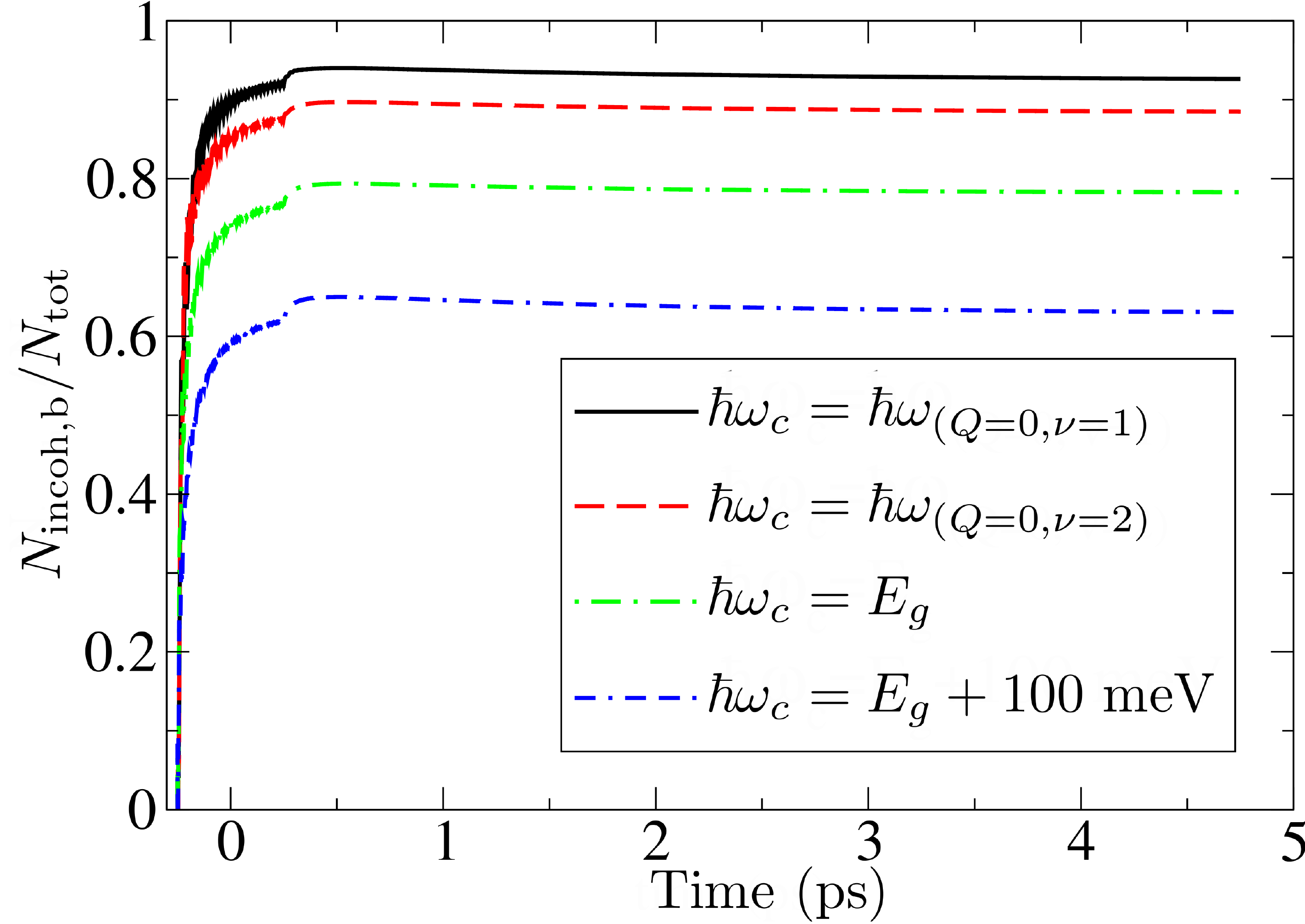}
 \end{center}
 \caption{(Color online) Time dependence of the relative number of incoherent bound excitons for different central frequencies of the pulse.}
 \label{fig:nu_0_diff_freq}
\end{figure}

We will firstly discuss the exciton formation process for different central frequencies of the exciting pulse.
We have considered central frequencies in resonance with $(Q=0,\nu=1)$ state,
$(Q=0,\nu=2)$ state, single--particle gap and
the central frequency which is $100\mrd\mathrm{meV}$ above the band gap.
As can be noted from Fig.~\ref{fig:nu_0_diff_freq}, raising the central frequency of the laser field leads to
lower relative number of incoherent bound excitons. Namely, the higher is the central frequency, the higher (in energy) are the
bands in which the initial coherent excitonic populations are created and the slower is the conversion of these coherent populations
to incoherent populations in lower excitonic bands. 
However, in the long--time limit, the relative number of incoherent bound excitons should not depend on the central frequency of the laser,
but tend to the value predicted by the Maxwell--Boltzmann distribution, which is above 99\%. Such a high value is due to the large energy
separation between the lowest excitonic band and the rest of the spectrum. We can thus infer, based on Fig.~\ref{fig:nu_0_diff_freq}, that
the semiconductor dynamics right after the pulsed excitation shows highly nonequilibrium features.
Relaxation towards equilibrium occurs on a time scale
longer than the picosecond one.

\begin{figure}[htbp]
 \begin{center}
  \includegraphics[width=0.4\textwidth]{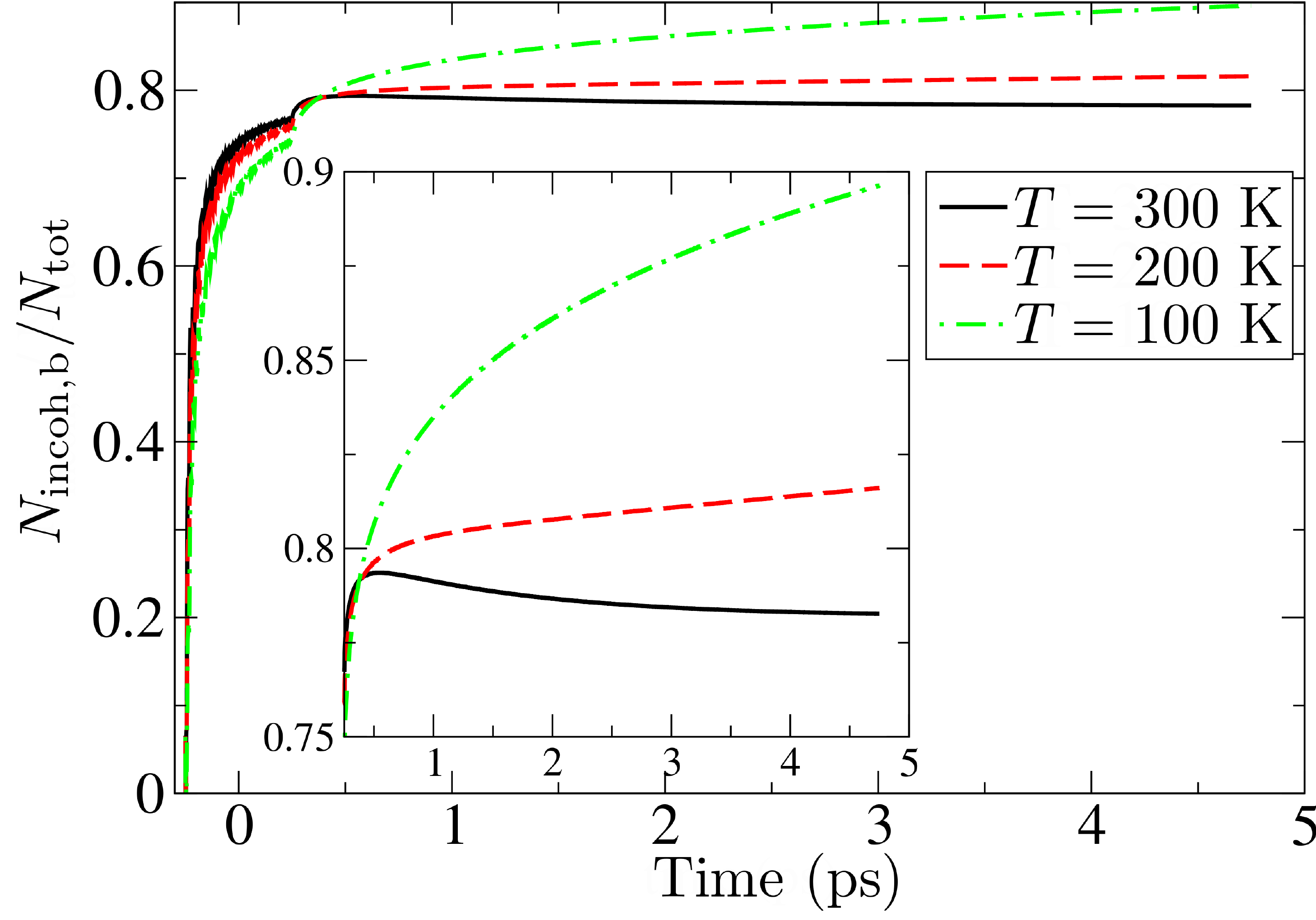}
 \end{center}
 \caption{(Color online) Time dependence of the relative number of incoherent bound excitons for different temperatures. The inset shows
the portions of the same curves after the pulse.}
 \label{fig:incoh_b_vs_T}
\end{figure}

\begin{figure}[htbp]
 \begin{center}
  \includegraphics[width=0.45\textwidth]{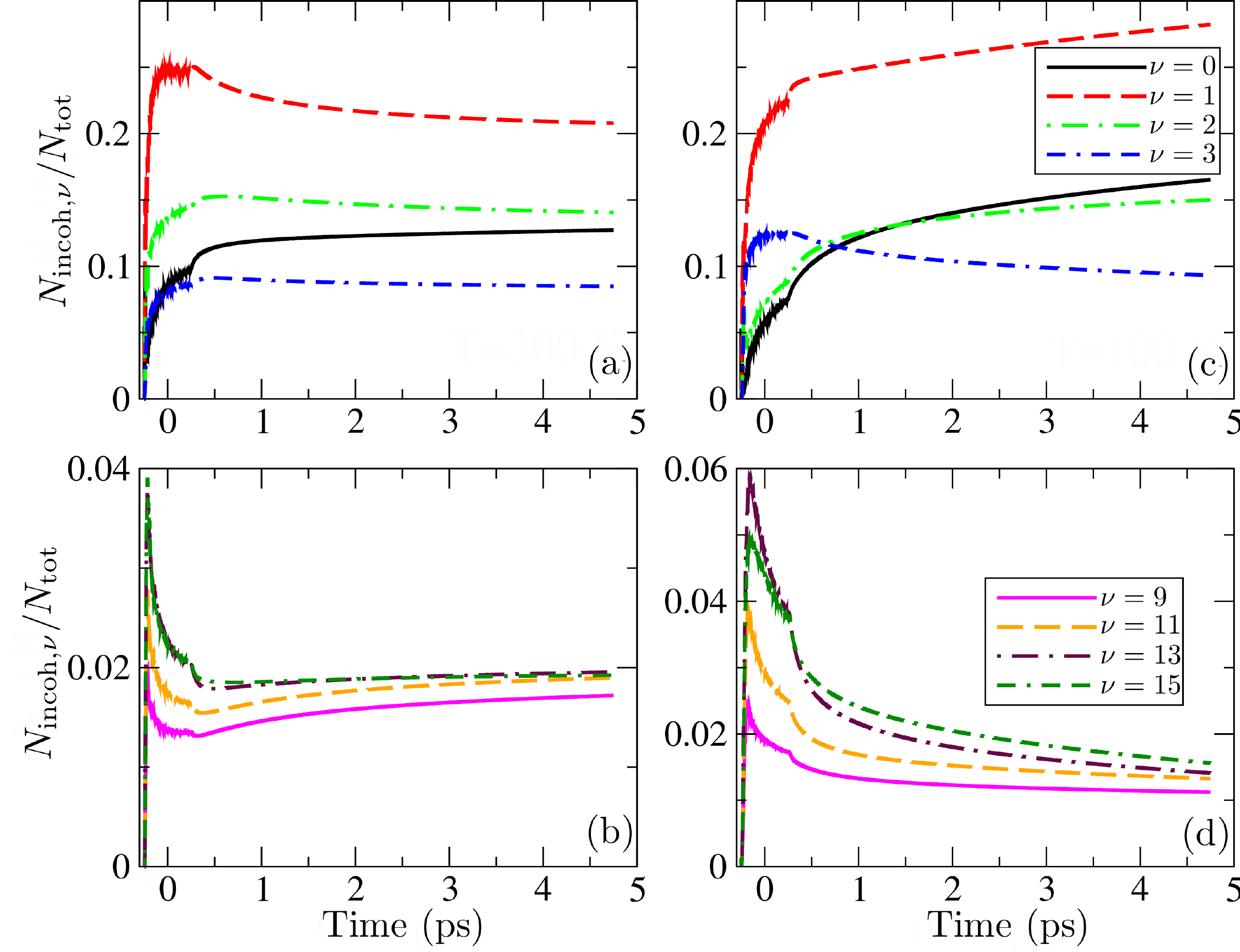}
 \end{center}
 \caption{(Color online) Time dependence of the relative population of various excitonic bands for different temperatures,
$T=300$ K for panels (a) and (b) and $T=100$ K for panels (c) and (d).
Panels (a) and (c) concern bands which are well below the pair continuum ($\nu=0,1,2,3$),
whereas panels (b) and (d) deal with the bands which are near the continuum
($\nu=9$) or in the continuum ($\nu=11,13,15$).}
\label{fig:org_T_bands}
\end{figure}

Next, we consider the dependence of the exciton formation process on temperature.
The temperature enters our model only through phonon numbers
$n^\mathrm{ph}$. The overall behavior of the relative number of incoherent bound excitons
for different temperatures
is shown in Fig.~\ref{fig:incoh_b_vs_T}.
During the pulse, the relative number of incoherent bound excitons is highest for $T=300$ K and lowest for $T=100$ K, which is
the consequence of the fact that scattering processes from higher excitonic bands
(in which initial coherent excitonic populations are created and which are situated both in the pair continuum and below it)
towards lower excitonic bands are most efficient at $T=300$ K. After the generation of carriers has been completed, phonon--mediated processes
lead to the redistribution of created incoherent excitons among different excitonic states and the relative number of incoherent bound excitons
increases with decreasing the temperature, which is the expected trend. In the inset of Fig.~\ref{fig:incoh_b_vs_T} we also note that the
relative number of incoherent bound excitons after the pulse experiences an initial growth followed by a slow decay at $T=300$ K, whereas
at $T=100$ K it monotonically rises. The initial growth at $T=300$ K is attributed to downward scattering processes, but since at this
temperature upward scattering events cannot be neglected, the following slow decay is due to the fact that
some excitonic bands well below the pair continuum (bands $\nu=1,2,3$) lose excitons
both by downward scattering and upward scattering to excitonic states which are near to or belong to the pair continuum
[see Figs.~\ref{fig:org_T_bands}(a) and~\ref{fig:org_T_bands}(b)].
At $T=100$ K, these upward processes are much less probable than downward processes,
thus the decay of the relative number of incoherent bound excitons is not observed;
in Figs.~\ref{fig:org_T_bands}(c) and~\ref{fig:org_T_bands}(d) we see that
lowest excitonic bands ($\nu=0,1,2$) gain excitons, whereas bands which are near to or belong to the pair continuum
($\nu=9,11,13,15$) lose excitons. The population of the lowest excitonic band $\nu=0$ continually grows at all the temperatures studied,
due to the large energetic separation between this band and the rest of the spectrum.

We briefly comment on the behavior of the number of coherent excitons $N_\mathrm{coh}$ and its temperature dependence.
\begin{figure}[htbp]
 \begin{center}
  \includegraphics[width=0.45\textwidth]{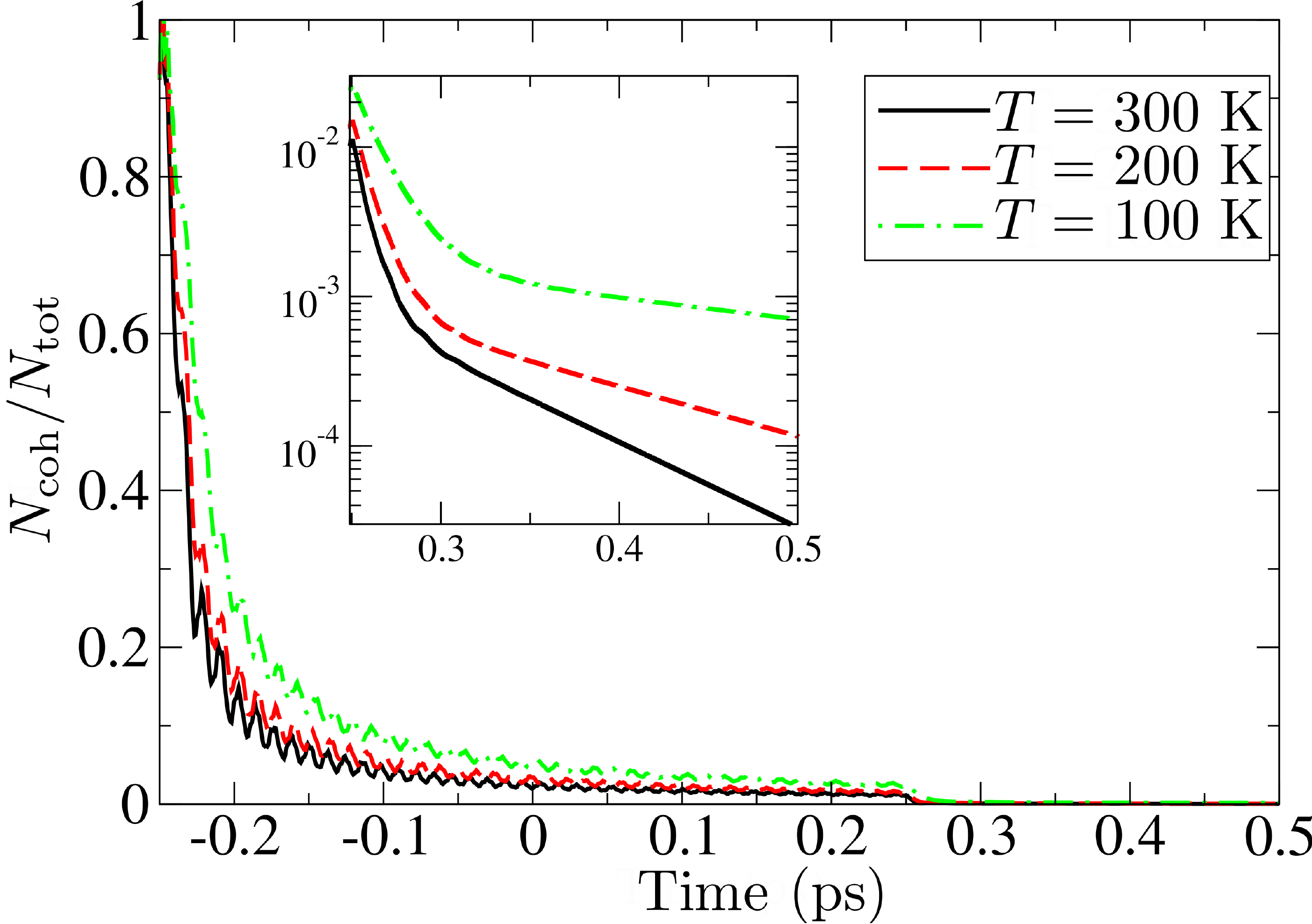}
 \end{center}
 \caption{(Color online) Time dependence of the relative number of coherent excitons for different temperatures. The inset shows the portions
of the same curves (note the logarithmic scale on the vertical axis) after the pulse.}
\label{fig:org_T_coh}
\end{figure}
Right after the start of the pulse, coherent excitons
comprise virtually the total excitonic population, see Fig.~\ref{fig:org_T_coh}.
Due to the carrier--phonon interaction, the relative number of coherent
excitons decays during the pulse, so that at its end coherent excitons comprise around 1\% of the total excitonic population.
The conversion from coherent to incoherent populations is thus almost completed by the end of the pulse.
From the inset of Fig.~\ref{fig:org_T_coh}, we note that $N_\mathrm{coh}/N_\mathrm{tot}$ exhibits a very fast decay after the pulse has vanished,
with decay times of the order of 50 fs or less.
Therefore, we infer that the transformation from coherent to incoherent excitonic populations takes place on a 50 fs time scale.
Based on Fig.~\ref{fig:org_T_coh}, we also note that the lower is the temperature,
the slower is the transformation from coherent to incoherent excitonic populations, which is the expected trend.

\begin{figure}[htbp]
 \begin{center}
  \includegraphics[width=0.4\textwidth]{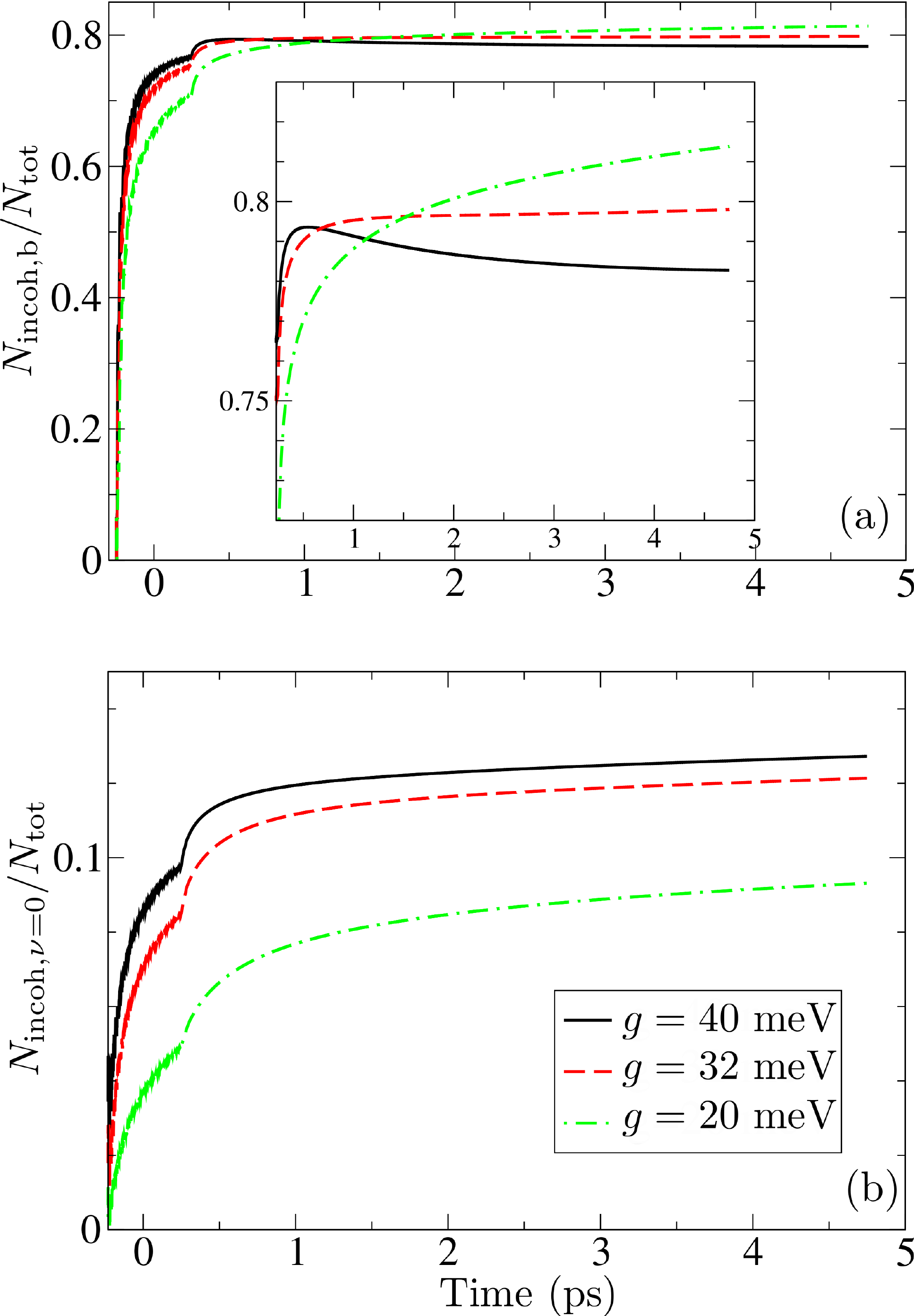}
 \end{center}
 \caption{(Color online) Time dependence of: (a) the relative number of incoherent bound excitons,
(b) the relative number of incoherent excitons in the $\nu=0$ band, for various values of $g$. The inset in the panel (a)
shows the portions of the same curves after the pulse.}
 \label{fig:incoh_b_vs_g}
\end{figure}

We continue our investigation by examining the effects that changes in the carrier--phonon coupling constant $g$ have on the exciton formation process.
Since increasing (lowering) $g$ increases (lowers) semiclassical transition rates,
just as increasing (lowering) $T$ does, the changes in $g$ and $T$ should have, in principle, similar effects on the exciton formation process.
Considering first
the relative number of incoherent bound excitons, whose time dependence for different values of $g$ is shown
in Fig.~\ref{fig:incoh_b_vs_g}(a), we note that after the end of the pulse it increases with decreasing $g$. However, during
the pulse, higher values of $g$ lead to more incoherent bound excitons, as is expected since scattering processes which populate
low--energy states are more intensive for larger $g$.
We also show the time dependence of the relative number of excitons in $\nu=0$ band in Fig.~\ref{fig:incoh_b_vs_g}(b).
It is observed that the lower is $g$, the lower is the number of excitons in the lowest excitonic band.
This is due to the fact that
populations on the lowest band are generated mainly via scattering processes
from the $\nu=1$ band and these processes are less efficient for smaller $g$.

\begin{figure}[htbp]
 \begin{center}
  \includegraphics[width=0.4\textwidth]{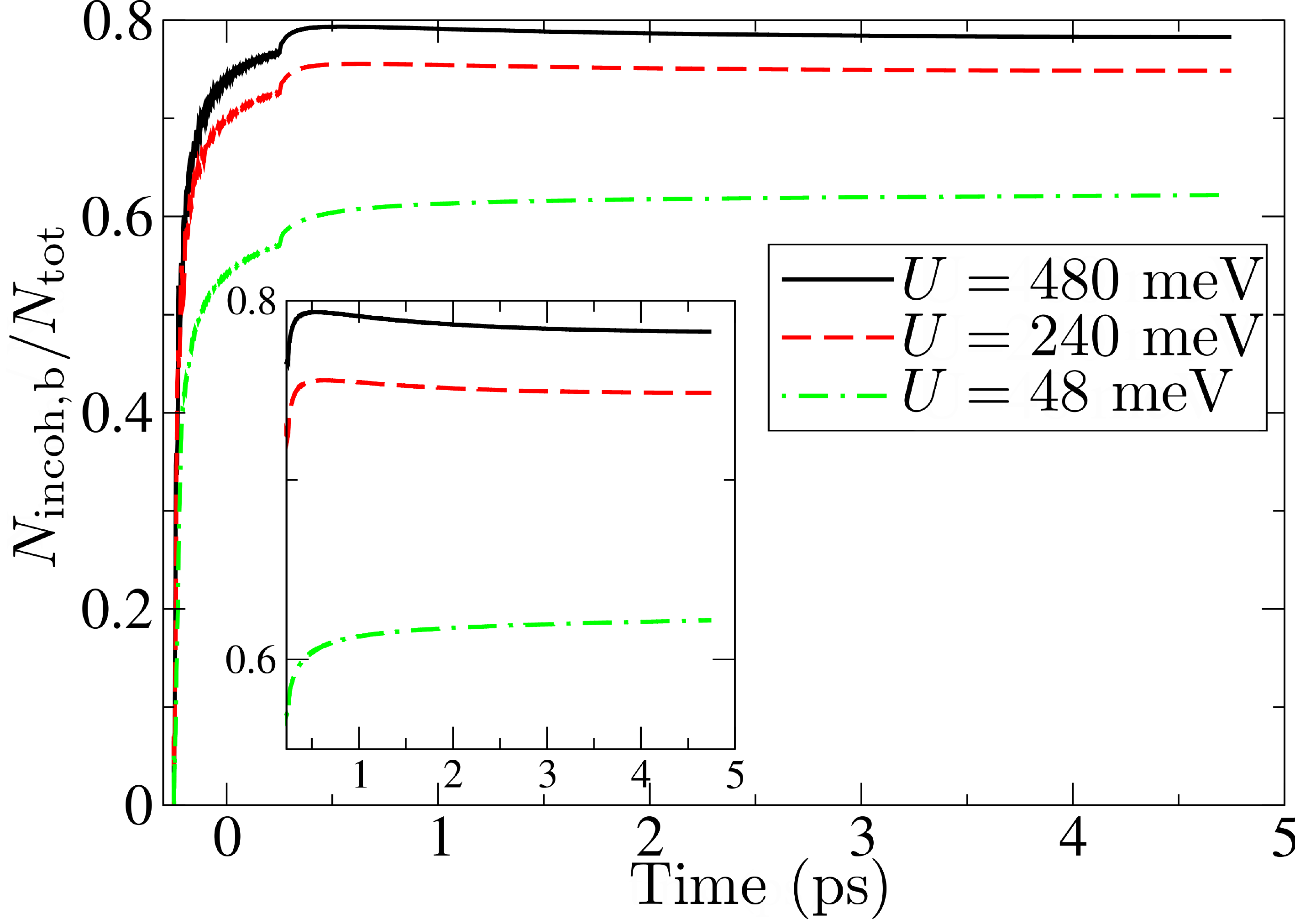}
 \end{center}
 \caption{(Color online) Time dependence of the relative number of incoherent bound excitons for various values of $U$.
The inset shows the portions of the same curves after the pulse.}
 \label{fig:incoh_b_vs_U}
\end{figure}

We conclude this section by studying the effects that changes in the on--site Coulomb interaction $U$ have on the process of exciton formation.
Changing $U$ has profound effects on the excitonic spectrum. Exciton binding energy lowers with lowering $U$ along with
the energy separation between the band $\nu=0$ and the rest of the spectrum. We studied the impact of $U$ on the exciton formation process
for three values of $U$, $U=480$ meV, $U=240$ meV and $U=48$ meV, for which the exciton binding energy is
$\sim 320$ meV, $\sim 175$ meV and $\sim 40$ meV, respectively.
Lowering $U$ lowers the relative number of incoherent bound excitons, as is shown
in Fig.~\ref{fig:incoh_b_vs_U}.
Smaller energy separation between the lowest excitonic band and the rest of the spectrum means that phonon--mediated transitions
which start/end on the band $\nu=0$ can end/start not predominantly on the band $\nu=1$, but also on higher excitonic bands,
which, for lower $U$, are more certain to belong to the electron--hole pair continuum than to the part of the spectrum which contains bound
pair states. Thus, the lower is $U$, the more likely are the dissociation processes in which an exciton, initially in a bound state,
after a phonon--mediated transition ends in an unbound pair state, which explains the observed trend in the relative number of incoherent
bound excitons. This agrees with the usual picture according to which thermal fluctuations are likely to dissociate loosely bound
electron--hole pairs. For $U=48$ meV, in the long--time limit and
according to the Maxwell--Boltzmann distribution, around $78\%$ of the total number of excitons should be in bound states, whereas
for the other two values of $U$ this number is above $99\%$. Thus, the dynamics observed is highly nonequilibrium, but unlike the cases
$U=480$ meV and $U=240$ meV, in which we cannot observe that the relative number of incoherent bound excitons starts to tend to its
equilibrium value, for $U=48$ meV we observe such a behavior (see the inset of Fig.~\ref{fig:incoh_b_vs_U}).

In summary, we list the time scales of the exciton formation and relaxation that stem from our computations. The transformation
from coherent to incoherent excitons takes place in less than 50 fs. Significant number of incoherent bound excitons is established on
a time scale of several hundreds of femtoseconds, whereas the subsequent relaxation of excitonic populations occurs on a time scale longer
than the picosecond one. Further discussion of these results is deferred for Sec.~\ref{Sec:discussion}.
\subsection{Numerical results: Inorganic set of parameters}
\begin{figure}[htbp]
 \begin{center}
  \includegraphics[width=0.4\textwidth]{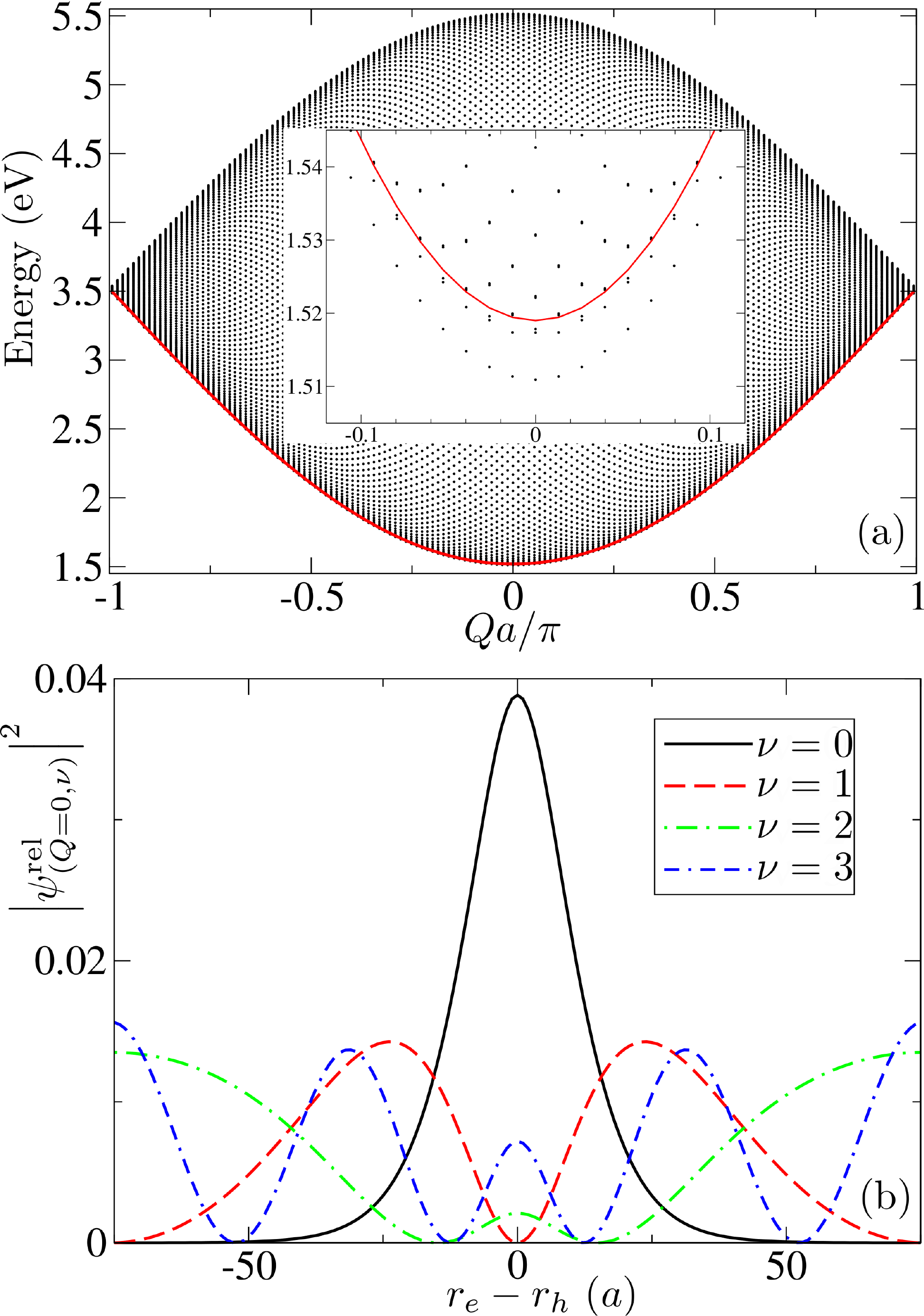}
 \end{center}
 \caption{(Color online)
(a) Excitonic spectrum for the inorganic set of parameters. Dots represent individual
excitonic states $(Q,\nu)$, while thick red line is the boundary between
bound and unbound excitonic states computed using Eq.~\eqref{Eq:sep}. The inset shows the same spectrum in the range of energies
around the single--particle gap.
(b) Squared modulus of the wave function which describes the relative motion of an electron--hole pair [Eq.~\eqref{Eq:ex_rel_mot}]
calculated for different states $(Q=0,\nu)$. Mean electron--hole
separations are $9.1a$ ($\nu=0$) and $29.4a$ ($\nu=1$), while states $(Q=0,\nu=2)$ and $(Q=0,\nu=3)$ are not bound.
Computations are performed for $N=151$.}
 \label{fig:spektar_inorg}
\end{figure}

In this section, we will investigate the exciton formation process in the case when material parameters assume values typical of
inorganic semiconductors, i.e., relatively large bandwidths, large dielectric constant (weak Coulomb interaction), and weak
carrier--phonon interaction. The excitonic spectrum is
shown in Fig.~\ref{fig:spektar_inorg}(a).
We see that almost all excitonic bands belong to the pair continuum, except for a couple of lowest bands, which is more clearly
seen in the inset of Fig.~\ref{fig:spektar_inorg}(a).
This is an entirely different situation from
the one that we encounter for the organic set of parameters,
where large energy separation of the lowest excitonic band from the
rest of the spectrum was crucial to understand the exciton formation process.
As a consequence, excitons in bound states are likely to scatter to a state in
the pair continuum, in contrast to the situation for the model parameters representative of an organic semiconductor.

Having noted the important characteristics of the excitonic spectrum, we move on to comment briefly on the exciton size
for the inorganic set of parameters. We plot in Fig.~\ref{fig:spektar_inorg}(b) the
squared modulus of the wave function of the relative motion of
the pair, which is defined in Eq.~\eqref{Eq:ex_rel_mot}.
We note that for the inorganic set of parameters, electron and hole are not as tightly bound as for the organic set
of parameters, which is in accord with the fact that excitons in a typical inorganic semiconductor have large radii,
typically of the order of 10 lattice constants.~\cite{Knoxbook,singhcontribution} From Fig.~\ref{fig:spektar_inorg}(b),
it is also clear that, if we are to
see the lowest excitonic state $(Q=0,\nu=0)$ as a bound pair, we should take the system size $N\gtrsim 120$.
We opted for $N=151$ because this value makes a good compromise between the minimal size of the system needed for
the results to be numerically accurate and the computational time.

\begin{figure}[htbp]
 \begin{center}
  \includegraphics[width=0.4\textwidth]{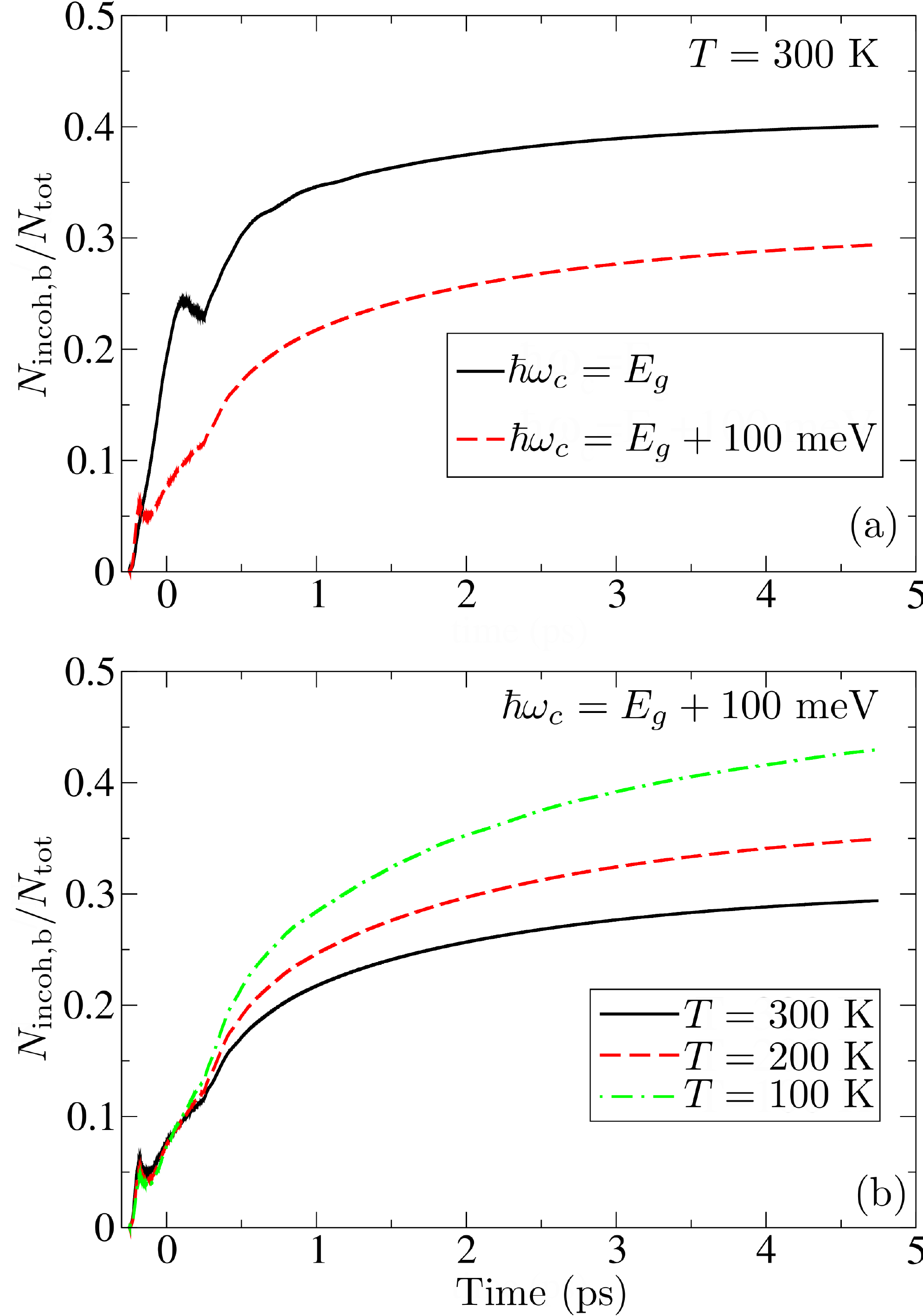}
 \end{center}
 \caption{(Color online) (a) Time dependence of the relative number of incoherent bound excitons for excitation resonant with the single--particle
gap and the one which is $100$ meV above it. The temperature in both cases is $T=300$ K.
(b) Time dependence of the relative number of incoherent bound excitons for various temperatures.
The central frequency of the laser pulse is $100$ meV above the single--particle gap.}
 \label{fig:inorg_incoh_b_vs_omegac}
\end{figure}

For the inorganic set of parameters, we note that incoherent unbound excitons comprise the major part of the total
excitonic population [see Fig.~\ref{fig:inorg_incoh_b_vs_omegac}(a)], which is
different from the case when model parameters assume values
representative of an organic semiconductor, when excitons in bound states prevail.
Considering an unbound exciton as quasifree
electron and hole, we interpret the last observation in the following manner:
after an optical excitation of an organic semiconductor,
(strongly) bound electron--hole pairs (excitons) are mainly generated,
whereas in the case of an inorganic semiconductor an optical excitation
predominantly generates (quasi) free charges.
In Fig.~\ref{fig:inorg_incoh_b_vs_omegac}, we also note that for higher central frequency of the laser field,
the relative number of bound excitons is lower. However, in the long--time limit the number of incoherent bound
excitons should assume the value predicted by the Maxwell--Boltzmann distribution, which is around 36.5\%, irrespectively
of the central frequency of the pulse. The values of the relative number of incoherent bound
excitons at the end of our computations do not strongly deviate from the value predicted by the Maxwell--Boltzmann distribution,
in contrast to the situation
for the organic set of parameters, where this deviation was more pronounced (see Fig.~\ref{fig:nu_0_diff_freq}). It can thus
be inferred that nonequilibrium features of the semiconductor dynamics after a pulsed excitation are more pronounced
for the organic than for the inorganic set of parameters.

Finally, we comment on the temperature dependence of the exciton formation process for the excitation whose central frequency
is $100$ meV above the single--particle gap. The lower is the temperature, the higher is the relative number of the incoherent
bound excitons [see Fig.~\ref{fig:inorg_incoh_b_vs_omegac}(b)]. During the pulse, higher temperature leads to higher relative number of
incoherent bound excitons, which has already been explained in the section dealing with the organic set of parameters.
The long--time limit values of the relative number of incoherent bound excitons are 44.7\% for $T=200$ K and 62.7\% for
$T=100$ K. In all three cases, the dynamics is highly nonequilibrium, but it displays the trend of a slow, but monotonic, approach
towards the equilibrium.

\section{Discussion}
\label{Sec:discussion}
In this section, we discuss the time scales of exciton formation and relaxation processes
obtained from our calculations in light of recent subpicosecond time--resolved experiments.
In Ref.~\onlinecite{jacs132-17459}, femtosecond--resolved fluorescence up--conversion spectroscopy was applied to investigate
the exciton dynamics in pristine PCDTBT polymer.
The results obtained were interpreted to originate from formation of free charges on less than 100 fs time scale, followed by formation
of bound excitons in less than 1 ps and their further relaxation at a longer time scale.
Similar results were obtained in Ref.~\onlinecite{jpcc115-9726} for P3HT polymer.
Despite the fact that our Hamiltonian does not include the effects of disorder that are present in real materials
and uses an oversimplified form of the carrier--phonon interaction,
we obtain time scales consistent with these data in our computations.
Namely, for the organic parameter set we find that significant population of bound excitons is formed on the time scale
of several hundreds of femtoseconds
and that their further relaxation occurs for at least several picoseconds.
These conclusions are further corroborated by fitting the relative number of incoherent bound excitons $N_{\mathrm{incoh,b}}/N_\mathrm{tot}$
after the carrier generation has been completed to a sum of three exponentially decaying terms. For the organic parameter set, we obtain characteristic time scales of
$\sim 50\mrd\mathrm{fs}$, $\sim 500\mrd\mathrm{fs}$ and $\gtrsim 1\mrd\mathrm{ps}$. We attribute the fastest time scale to decoherence
processes which are responsible for conversion from coherent ($|y_x|^2$) to incoherent ($\bar n_{xx}$)
populations due to the interaction with phonons. The time scale
of $\sim 500\mrd\mathrm{fs}$ may be associated with the build--up of the Coulomb--induced correlations between electrons and holes by
formation of bound incoherent electron--hole pairs via phonon--assisted
scattering processes. After this time scale, however, intraband coherences
$\bar n_{\bar x x}$ ($\bar x\neq x$), as well as single--phonon--assisted
density matrices $n_{\bar x x\mu^+}$, still have significant values. In the long--time limit, these variables asymptotically
vanish, and we remain only with incoherent populations whose dynamics will eventually lead to
thermalized distribution of excitons.~\cite{RevModPhys.70.145}
As our computations are certainly not long enough to observe these effects, we speculate
that the slowest time scale we obtain may be related to the decay of the intraband coherences and/or phonon--assisted variables.

Next, we comment on the relation of our results with recent experimental insights which have challenged the commonly accepted
physical picture of the generation of free charges in bulk heterojunction solar cells. Namely, it is widely believed that physical processes leading to
current generation are formation of bound excitons due to light absorption in the donor material, their
diffusion to the donor/acceptor interface, and their subsequent separation at the interface.~\cite{ChemRev.110.6736}
From the discrepancy between the distance that a donor exciton can diffuse in $100\mrd\mathrm{fs}$ and the distance it has to cover in order to
reach the donor/acceptor interface in efficient bulk heterojunction solar cells, Cowan {\it et al.}~\cite{afm22-1116} conclude that the subpicosecond
charge transfer to the acceptor occurs before exciton formation in the donor. The results of our computations,
which indicate that the formation of incoherent bound
excitons occurs on a $\sim 500\mrd\mathrm{fs}$ time scale, are therefore consistent with their observations.
The formation of hot charge transfer excitons which occurs in less than 100 fs and which is followed by their relaxation to
lower energies and shorter electron--hole distances on a picosecond time scale was experimentally observed in a small molecule CuPc/fullerene
blend using time--resolved second harmonic generation
and time--resolved two--photon photoemission.~\cite{nmat12-66}
The presence of hot charge transfer excitons, which are delocalized, i.e., in which the electron--hole separation is rather large, and their essential
role in subpicosecond charge separation in efficient OPV systems were also identified in Ref.~\onlinecite{nmat12-29,science335-1340,science343-512}.
Our simulation results that indicate exciton equilibration times longer than picoseconds are fully consistent with observations that during charge
separation at the donor/acceptor interface the excitons remain out of equilibrium (hot excitons).

\section{Conclusion}
\label{Sec:conclusion}
In conclusion, we have investigated the exciton dynamics in a photoexcited semiconductor on a picosecond time scale.
The study was conducted on the two--band semiconductor Hamiltonian, which includes relevant physical effects in the system,
using the density matrix theory combined with the DCT scheme. We truncate the phonon branch of the hierarchy and propose the form
of coupling between electronic density matrices with single--phonon assistance and higher--order phonon assistance
so as to achieve the compatibility of the resulting equations with the energy and particle--number conservation in a system without external fields.
The numerical study aiming at identifying time scales of exciton formation and relaxation processes
was performed on a one--dimensional model system for the values of model parameters representative of
a typical organic and inorganic semiconductor.
We concluded that the dynamics on a picosecond time scale shows highly nonequilibrium features,
relaxation processes towards equilibrium occurring on a longer time scale.
While for the organic set of parameters the excitons generated are mainly tightly bound, for the inorganic set of
parameters the major part of excitons is in unbound pair states and may thus be considered as
(quasi)free electrons and holes. In other words, a photoexcitation of an initially
unexcited organic semiconductor leads to creation of bound electron--hole pairs, whereas in an inorganic semiconductor
it leads to generation of free charges.
This difference can be mainly attributed to different properties of the excitonic spectrum,
which for the organic set of parameters exhibits large energy separation between the lowest excitonic band and the rest
of the spectrum.
Furthermore, although the carrier--phonon interaction is stronger for the organic set of parameters,
we have noted that the number of excitons in bound states more strongly deviates from its equilibrium value for the organic
set of parameters than for the inorganic one. This observation emphasizes the importance of nonequilibrium effects for the proper
understanding of the ultrafast dynamics of photoexcited organic semiconductors and unraveling the working principles of organic photovoltaic devices.    
\acknowledgments
We gratefully acknowledge the support by Serbian Ministry of Education, Science and Technological
Development (Project No. ON171017) and European Community FP7
Marie Curie Career Integration Grant (ELECTROMAT) and the contribution of the COST Action MP1406. Numerical simulations were performed on the PARADOX supercomputing
facility at the Scientific Computing Laboratory of the Institute of Physics Belgrade.

\begin{widetext}
\appendix
\section{Equations of motion}\label{App:eom}
In this appendix, we present equations of motion for relevant dynamic variables. These are the same equations as
in Ref.~\onlinecite{RevModPhys.70.145}, with only slight modifications in notation,
which are exact up to the second order in the external field. We point out that, according to the generating
function property, differential equations for the corresponding phonon--assisted density matrices are obtained after
performing appropriate differentiations and setting $\alpha_\mu=\beta_\mu=0$:
\begin{equation}
\label{Eq:eom_y}
\begin{split}
 \im\hbar\mrd\partial_t Y_{ab}^{\alpha\beta}&=
 (\epsilon^\cb_b-\epsilon^\vb_a)Y_{ab}^{\alpha\beta}+\sum_{\substack{p\in\mathrm{VB}\\q\in\mathrm{CB}}}
 \left(V^{\vb\cb\cb\vb}_{pqba}-V^{\vb\vb\cb\cb}_{pabq}\right)Y_{pq}^{\alpha\beta}
 +\sum_{\mu}\hbar\omega_\mu\left(\beta_\mu\partial_{\beta_\mu}-\alpha_\mu\partial_{\alpha_\mu}\right)Y_{ab}^{\alpha\beta}\\
 &+\sum_{\substack{k\in\mathrm{CB}\\\mu}}
 \left(\gamma^\mu_{bk}(\partial_{\alpha_\mu}+\beta_\mu)+\gamma^{\mu*}_{kb}\partial_{\beta_\mu}\right)Y_{ak}^{\alpha\beta}
 -\sum_{\substack{k\in\mathrm{VB}\\\mu}}
 \left(\gamma^\mu_{ka}(\partial_{\alpha_\mu}+\beta_\mu)+\gamma^{\mu*}_{ak}\partial_{\beta_\mu}\right)Y_{kb}^{\alpha\beta}\\
 &-\mathbf{E}(t)\mathbf{M}^{\cb\vb}_{ba}F^{\alpha\beta},
 \end{split}
\end{equation}
\begin{equation}
\label{Eq:eom_n}
 \begin{split}
  \im\hbar\mrd\partial_t N_{abcd}^{\alpha\beta}&=(\epsilon^\cb_d-\epsilon^\vb_c+\epsilon^\vb_b-\epsilon^\cb_a)N_{abcd}^{\alpha\beta}
  +\sum_{\substack{p\in\mathrm{VB}\\q\in\mathrm{CB}}}\left(
  \left(V^{\vb\cb\cb\vb}_{pqdc}-V^{\vb\vb\cb\cb}_{pcdq}\right)N_{abpq}^{\alpha\beta}-
  \left(V^{\vb\cb\cb\vb}_{baqp}-V^{\vb\vb\cb\cb}_{bpqa}\right)N_{qpcd}^{\alpha\beta}\right) \\
  &+\sum_{\mu}\hbar\omega_\mu\left(\beta_\mu\partial_{\beta_\mu}-\alpha_\mu\partial_{\alpha_\mu}\right)N_{abcd}^{\alpha\beta}\\
  &+\sum_{\substack{k\in\mathrm{CB}\\\mu}}\left(
  \left(\gamma^\mu_{dk}(\partial_{\alpha_\mu}+\beta_\mu)+\gamma^{\mu*}_{kd}\partial_{\beta_\mu}\right)N_{abck}^{\alpha\beta}
  -\left(\gamma^\mu_{ka}\partial_{\alpha_\mu}+\gamma^{\mu*}_{ak}(\partial_{\beta_\mu}+\alpha_\mu)\right)N_{kbcd}^{\alpha\beta}
  \right)\\
  &-\sum_{\substack{k\in\mathrm{VB}\\\mu}}\left(
  \left(\gamma^\mu_{kc}(\partial_{\alpha_\mu}+\beta_\mu)+\gamma^{\mu*}_{ck}\partial_{\beta_\mu}\right)N_{abkd}^{\alpha\beta}
  -\left(\gamma^\mu_{bk}\partial_{\alpha_\mu}+\gamma^{\mu*}_{kb}(\partial_{\beta_\mu}+\alpha_\mu)\right)N_{akcd}^{\alpha\beta}
  \right)\\
  &-\mathbf{E}(t)\left(\mathbf{M}^{\cb\vb}_{dc}Y_{ba}^{\beta\alpha*}-
  \mathbf{M}^{\vb\cb}_{ba}Y_{cd}^{\alpha\beta}\right).
 \end{split}
\end{equation}
\section{Closing the hierarchy of equations}\label{App:two-ph}
In Eq.~\eqref{Eq:dif_n_barx_x_mu_piu}, correlated parts of two--phonon--assisted density matrices $\delta n_{\bar x x\rho^{+}\sigma^{-}}$
and $\delta n_{\bar x x\rho^+\sigma^+}$ appear. In their differential equations, three--phonon--assisted density matrices are present. In
order to close the hierarchy of equations, we factorize them into all possible combinations of phonon distribution functions and phonon--assisted
electronic density matrices and neglect their correlated parts. The strategy for the factorization is
the one we employed in Eq.~\eqref{Eq:n_two_ph_ass_fact} where we considered an exciton as a basic entity and did not take into account
contributions arising from the excitonic amplitude (with possible phonon assistance).
Namely, the two--phonon--assisted electronic density matrix $\langle c_a^\dagger d_b^\dagger d_c c_d b_\mu^\dagger b_\rho\rangle$
can be written in terms of exciton creation and
annihilation operators [see Eq.~\eqref{Eq:def_exc_creat}] as
$\displaystyle{\sum_{\bar x x}\psi^{\bar x*}_{ba}\psi^x_{cd}\langle X_{\bar x}^\dagger X_{x} b_\mu^\dagger b_\rho\rangle}$.
Since it appears in the equation of motion for one--phonon--assisted electronic density matrix $n_{\bar x x\mu}^{(+)}$, which is
coupled to Eq.~\eqref{Eq:dif_n_barx_x} describing excitonic populations and intraband coherences, we treat an exciton as a basic entity
and accordingly perform the factorization
$\displaystyle{
\langle X_{\bar x}^\dagger X_{x} b_\mu^\dagger b_\rho\rangle=\langle X_{\bar x}^\dagger X_{x}\rangle\langle b_\mu^\dagger b_\rho\rangle+
\delta\langle X_{\bar x}^\dagger X_{x} b_\mu^\dagger b_\rho\rangle
}.$ 
In the case of three--phonon--assisted electronic density matrices, the described factorization procedure, neglecting the correlated part, gives
\begin{equation}
 \langle c_a^\dagger d_b^\dagger d_c c_d b_\mu^\dagger b_\rho^\dagger b_\sigma\rangle=
\delta_{\rho\sigma}\langle c_a^\dagger d_b^\dagger d_c c_d b_\mu^\dagger\rangle n_\rho^\mathrm{ph}+
\delta_{\mu\sigma}\langle c_a^\dagger d_b^\dagger d_c c_d b_\rho^\dagger\rangle n_\mu^\mathrm{ph}.
\end{equation}
Performing transition to the excitonic basis, the following differential equation for the variable $\delta n_{\bar x x\rho^+\sigma^-}$ is obtained:
\begin{equation}
\label{Eq:dif_n_2_ph_ass}
\begin{split}
 \partial_t\mrd\delta n_{\bar x x\rho^+\sigma^-}&=-\im(\omega_x-\omega_{\bar x}+\omega_\sigma-\omega_\rho)\delta n_{\bar x x\rho^+\sigma^-}\\
&+\frac{1+n_\sigma^\mathrm{ph}}{\im\hbar}\sum_{x'}\Gamma^\sigma_{xx'} n_{\bar x x'\rho^+}
-\frac{n_\sigma^\mathrm{ph}}{\im\hbar}\sum_{\bar x'}\Gamma^\sigma_{\bar x'\bar x} n_{\bar x' x \rho^+}\\
&-\frac{1+n_\rho^\mathrm{ph}}{\im\hbar}\sum_{\bar x'}\Gamma^{\rho*}_{\bar x\bar x'} n_{x \bar x'\sigma^+}^{*}
+\frac{n_\rho^\mathrm{ph}}{\im\hbar}\sum_{x'}\Gamma^{\rho*}_{x'x} n_{x' \bar x\sigma^+}^{*},
\end{split}
\end{equation}
and similarly for the variable $\delta n_{\bar x x\rho^+\sigma^+}$.
Solving Eq.~\eqref{Eq:dif_n_2_ph_ass} in the Markov and adiabatic approximations,~\cite{PhysRevB.50.5435,PhysRevB.46.7496} the following result is obtained
\begin{equation}
\label{Eq:sol_Mar_adi}
 \begin{split}
 \delta n_{\bar x x\rho^+\sigma^-}&=
 (1+n_\sigma^\mathrm{ph})\sum_{x'}\Gamma^\sigma_{xx'}\mathcal{D}(\hbar\omega_{x'}-\hbar\omega_x-\hbar\omega_\sigma)n_{\bar x x'\rho^+}
 -n_\sigma^\mathrm{ph}\sum_{\bar x'}\Gamma^\sigma_{\bar x'\bar x}\mathcal{D}(\hbar\omega_{\bar x}-\hbar\omega_{\bar x'}-\hbar\omega_\sigma)n_{\bar x' x \rho^+}\\
 &+
 (1+n_\rho^\mathrm{ph})\sum_{\bar x'}\Gamma^{\rho*}_{\bar x\bar x'}\mathcal{D}^*
 (\hbar\omega_{\bar x'}-\hbar\omega_{\bar x}-\hbar\omega_\rho)n_{x \bar x'\sigma^+}^{*}
 -n_\rho^\mathrm{ph}
 \sum_{x'}\Gamma^{\rho*}_{x'x}
 \mathcal{D}^*(\hbar\omega_x-\hbar\omega_{x'}-\hbar\omega_\rho)n_{x' \bar x\sigma^+}^{*},
 \end{split}
\end{equation}
where $\mathcal{D}(\epsilon)=-\im\pi\delta(\epsilon)+\mathcal{P}(1/\epsilon)$.
We thus expressed two--phonon--assisted electronic density matrices in terms of one--phonon--assisted
electronic density matrices.
When these results are inserted
in Eq.~\eqref{Eq:dif_n_barx_x_mu_piu}, we neglect all terms involving principal values which, in principle,
lead to polaron shifts in energies.~\cite{PhysRevB.50.5435,kuhncontribution}
Furthermore, we note that
the inserted terms involve multiple summations over excitonic indices $x$ and we use the random phase approximation
to simplify the expression obtained.
This approximation is easier to understand and justify when we transfer to a particular representation for the excitonic
index $x$, for example, the one that we used in our computational study, where we took advantage of the translational symmetry and
had $x=(Q,\nu)$. 
Electronic density matrices with one--phonon assistance $n_{(\bar Q,\bar\nu)(Q,\nu)q_\mu^+}$
are complex quantities, which acquire nontrivial values during the evolution provided that the condition $\bar Q+q_\mu=Q$ is satisfied.
Having in mind the selection rule for carrier--phonon matrix elements in the excitonic basis [see Eq.~\eqref{Eq:carr_phon_sel_r}],
we can express the first term which describes the coupling of the one--phonon--assisted electronic density matrix
$n_{(Q-q_\mu,\bar\nu)(Q,\nu)q_\mu^+}$ to density matrices with higher phonon assistance [see Eq.~\eqref{Eq:dif_n_barx_x_mu_piu}] as
\begin{equation}
 \begin{split}
  -&\frac{1}{\im\hbar}\sum_{\rho \bar x'}\Gamma^{\rho*}_{\bar x\bar x'}\delta n_{\bar x' x\mu^+\rho^-}\\
&\frac{\pi}{\hbar}\sum_{q_\rho,\nu',\bar\nu'}
\Gamma^{q_\rho*}_{(Q-q_\mu,\bar\nu)(Q-q_\mu+q_\rho,\bar\nu')}
\Gamma^{q_\rho}_{(Q,\nu)(Q+q_\rho,\nu')}
(1+n_{q_\rho}^\mathrm{ph})
\delta(\hbar\omega_{(Q+q_\rho,\nu')}-\hbar\omega_{(Q,\nu)}-\hbar\omega_{q_\rho})
n_{(Q-q_\mu+q_\rho,\bar\nu')(Q+q_\rho,\nu')q_\mu^+}\\
-&\frac{\pi}{\hbar}\sum_{q_\rho,\bar\nu',\bar\nu''}
\Gamma^{q_\rho*}_{(Q-q_\mu,\bar\nu)(Q-q_\mu+q_\rho,\bar\nu')}
\Gamma^{q_\rho}_{(Q-q_\mu,\bar\nu'')(Q-q_\mu+q_\rho,\bar\nu')}
n_{q_\rho}^\mathrm{ph}
\delta(\hbar\omega_{(Q-q_\mu+q_\rho,\bar\nu')}-\hbar\omega_{(Q-q_\mu,\bar\nu'')}-\hbar\omega_{q_\rho})
n_{(Q-q_\mu,\bar\nu'')(Q\nu)q_\mu^+}\\
-&\frac{\pi}{\hbar}\sum_{q_\rho,\bar\nu',\bar\nu''}
\Gamma^{q_\rho*}_{(Q-q_\mu,\bar\nu)(Q-q_\mu+q_\rho,\bar\nu')}
\Gamma^{q_\mu*}_{(Q-q_\mu+q_\rho,\bar\nu')(Q+q_\rho,\bar\nu'')}
(1+n_{q_\mu}^\mathrm{ph})
\delta(\hbar\omega_{(Q+q_\rho,\bar\nu'')}-\hbar\omega_{(Q-q_\mu+q_\rho,\bar\nu')}-\hbar\omega_{q_\mu})
n_{(Q,\nu)(Q+q_\rho,\bar\nu'')q_\rho^+}^*\\
+&\frac{\pi}{\hbar}\sum_{q_\rho,\nu',\bar\nu'}
\Gamma^{q_\rho*}_{(Q-q_\mu,\bar\nu)(Q-q_\mu+q_\rho,\bar\nu')}
\Gamma^{q_\mu*}_{(Q-q_\mu,\nu')(Q,\nu)}
n_{q_\mu}^\mathrm{ph}
\delta(\hbar\omega_{(Q,\nu)}-\hbar\omega_{(Q-q_\mu,\nu')}-\hbar\omega_{q_\mu})
n_{(Q-q_\mu,\bar\nu')(Q-q_\mu+q_\rho,\nu')q_\rho^+}^*
 \end{split}
\end{equation}
 
In the first, the third, and the fourth sums in the previous equation we perform summation of terms which involve
complex--valued single--phonon--assisted electronic density matrices
over the wave vector $q_\rho$, whereas in the second sum the summation is not carried out over any of the wave vectors
describing the density matrix. In the lowest approximation, we can assume that all the sums apart from the second are negligible
due to random phases at different wave vectors. For the sake of simplicity, in the second sum we keep only the contribution
for $\bar\nu''=\bar\nu$, thus expressing the coupling to higher--phonon--assisted density matrices only in terms of the
single--phonon--assisted density matrix for which the equation is formed. Restoring the more general notation, we obtain the result
\begin{equation}
\label{Eq:first_term_RPA}
 -\frac{1}{\im\hbar}\sum_{\rho \bar x'}\Gamma^{\rho*}_{\bar x\bar x'}\delta n_{\bar x' x\mu^+\rho^-}=
-\frac{\pi}{\hbar}\left(\sum_{\rho\tilde x}|\Gamma^\rho_{\bar x\tilde x}|^2
n_\rho^\mathrm{ph}\delta(\hbar\omega_{\bar x}-\hbar\omega_{\tilde x}+\hbar\omega_\rho)\right)
n_{\bar x x\mu^+}.
\end{equation}
Repeating similar procedure with the remaining three terms which describe coupling to density matrices with higher--order phonon
assistance in Eq.~\eqref{Eq:dif_n_barx_x_mu_piu}, we obtain the result embodied in
Eqs.~\eqref{Eq:result_RPA}$-$\eqref{Eq:attach_def_gamma_barx_x_mu}. 

Analogously,
the following results
for two--phonon--assisted electronic density matrices $\delta y_{x\rho^+\sigma^-},\delta y_{x\rho^+\sigma^+}$
are obtained, solving their respective differential equations in the Markov and adiabatic approximations
\begin{equation}
 \delta y_{x\rho^+\sigma^-}=(1+n_\sigma^\mathrm{ph})\sum_{x'}
 \Gamma^\sigma_{xx'}\mathcal{D}(\hbar\omega_{x'}-\hbar\omega_x-\hbar\omega_\sigma) y_{x'\rho}^{(+)}
 -n_\rho^\mathrm{ph}\sum_{x'}
 \Gamma^{\rho*}_{x'x}\mathcal{D}^*(\hbar\omega_x-\hbar\omega_{x'}-\hbar\omega_\rho) y_{x'\sigma}^{(-)},
\end{equation}
and similarly for the variable $\delta y_{x\rho^+\sigma^+}$. Inserting the results obtained in Eqs.~\eqref{Eq:dif_y_x_mu_piu}
and~\eqref{Eq:dif_y_x_mu_meno} and performing the random phase approximation as described, the result given
in Eqs.~\eqref{Eq:result_apps_y} and~\eqref{Eq:result_apps_y_c} is obtained.
\section{Comments on the energy conservation in the model}~\label{App:ene_con}
In this appendix, we will comment on the energy conservation in the model after the external field has vanished. Using
Eqs.~\eqref{Eq:carr_ene},~\eqref{Eq:phon_ene},~\eqref{Eq:dif_n_barx_x}, and~\eqref{Eq:dif_num_ph},
we obtain the rate at which the energy of carriers and phonons changes after the pulse
\begin{equation}
 \partial_t\mrd(\mathcal{E}_\cb+\mathcal{E}_\mathrm{ph})=
-\frac{2}{\hbar}\sum_{\mu\bar x x}(\hbar\omega_x-\hbar\omega_{\bar x}-\hbar\omega_\mu)
\mathrm{Im}\{\Gamma^\mu_{\bar x x}n_{\bar x x\mu^+}\},
\end{equation}
which exactly cancels the part from $\partial_t\mrd\mathcal{E}_{\mathrm{c}-\mathrm{ph}}$
[see Eq.~\eqref{Eq:carr_phon_ene}]
that originates from the free rotation term $-\im(\omega_x-\omega_{\bar x}-\omega_\mu)n_{\bar x x \mu^+}$
in Eq.~\eqref{Eq:dif_n_barx_x_mu_piu}. The terms in $\partial_t\mrd\mathcal{E}_{\mathrm{c}-\mathrm{ph}}$
which arise from the second and third terms in Eq.~\eqref{Eq:dif_n_barx_x_mu_piu} are
identically equal to zero each since they are purely real, which is easily
checked. Therefore, the rate at which the total energy changes after the pulse is equal to
the rate at which the carrier--phonon interaction energy changes due to the coupling of single--phonon--assisted
to higher--order phonon--assisted density matrices,
$\left(\partial_t\mrd\mathcal{E}_\mathrm{c-ph}\right)_\mathrm{higher}$, which is equal to [see Eq.~\eqref{Eq:dif_n_barx_x_mu_piu}]
\begin{equation}
\label{Eq:dif_e_c_ph_higher}
\begin{split}
  \left(\partial_t\mrd\mathcal{E}_\mathrm{c-ph}\right)_\mathrm{higher}=
&-\frac{2}{\hbar}\mathrm{Im}\left\{
\sum_{\substack{\mu\bar x x\\\rho\bar x'}}\Gamma^\mu_{\bar x x}\Gamma^{\rho*}_{\bar x\bar x'}\delta n_{\bar x' x\mu^+\rho^-}
\right\}
-\frac{2}{\hbar}\mathrm{Im}\left\{
\sum_{\substack{\mu\bar x x\\\rho\bar x'}}\Gamma^\mu_{\bar x x}\Gamma^{\rho}_{\bar x'\bar x}\delta n_{\bar x' x\mu^+\rho^+}
\right\}\\
&+\frac{2}{\hbar}\mathrm{Im}\left\{
\sum_{\substack{\mu\bar x x\\\rho x'}}\Gamma^\mu_{\bar x x}\Gamma^{\rho*}_{x'x}\delta n_{\bar x x'\mu^+\rho^-}
\right\}
+\frac{2}{\hbar}\mathrm{Im}\left\{
\sum_{\substack{\mu\bar x x\\\rho x'}}\Gamma^\mu_{\bar x x}\Gamma^{\rho}_{xx'}\delta n_{\bar x x'\mu^+\rho^+}
\right\}
\end{split}
\end{equation}
The first and the third terms on the right--hand side of Eq.~\eqref{Eq:dif_e_c_ph_higher} are separately equal to zero
(since the quantities under the sign of the imaginary part are purely real), whereas the second and the fourth terms exactly cancel
each other, so the total energy is conserved. In particular, this is true for the form of the correlated parts of two--phonon--assisted
density matrix $\delta n_{\bar x x\rho^+\sigma^-}$ given in Eq.~\eqref{Eq:sol_Mar_adi} and the similar form
of the density matrix $\delta n_{\bar x x\rho^+\sigma^+}$.
In Eq.~\eqref{Eq:dif_e_c_ph_higher}, all the sums are performed over all indices that
are present in a particular expression, so the crux of the proof that the energy is conserved is the interchange of dummy indices
combined with the properties $\delta n_{\bar x x\rho^+\sigma^-}^{*}=\delta n_{x \bar x\sigma^+\rho^-}$ and
$\delta n_{\bar x x\rho^+\sigma^+}=\delta n_{\bar x x\sigma^+\rho^+}$. However, when we apply the random phase approximation,
the aforementioned properties are lost and the energy is not conserved any more. For example, the first term on the right--hand side
in Eq.~\eqref{Eq:dif_e_c_ph_higher} after performing the random phase approximation is not equal to zero, but to
$\displaystyle{-\frac{2\pi}{\hbar}\left(\sum_{\rho\tilde x}|\Gamma^\rho_{\bar x\tilde x}|^2
n_\rho^\mathrm{ph}\delta(\hbar\omega_{\tilde x}-\hbar\omega_{\bar x}+\hbar\omega_\rho)\right)
\mathrm{Re}\left\{\sum_{\mu\bar x x}\Gamma^\mu_{\bar x x} n_{\bar x x\mu^+}\right\}}$ [see Eq.~\eqref{Eq:first_term_RPA}], which is
just one term of the total rate $\left(\partial_t\mrd\mathcal{E}_\mathrm{c-ph}\right)_\mathrm{higher}$ when we use the result from
Eq.~\eqref{Eq:result_RPA}.
\end{widetext}

\newpage

\bibliography{refs}

\end{document}